\documentclass[11pt,a4paper]{article}\pdfoutput=1

\usepackage{jheppub}
\usepackage{hyperref}
\usepackage{booktabs}
\usepackage{xspace}
\usepackage[T1]{fontenc}
\usepackage{xcolor}
\usepackage[toc,page]{appendix}
\usepackage{mathtools}
\usepackage[utf8]{inputenc}

\usepackage{slashed}

\newcommand{\noun}[1]{{\tt #1}}
\newcommand{\sss}{\scriptscriptstyle }
\newcommand{\MSbar}{\overline{\rm MS}}
\newcommand{\POWHEG}{\noun{POWHEG}}
\newcommand{\LHAPDF}{\noun{LHAPDF}}
\newcommand{\POWHEGHOOK}{\noun{PowhegHooks}}

\newcommand{\POWHEGBOXRES}{\noun{POWHEG-BOX-RES}}
\newcommand{\POWHEGBOXVTWO}{\noun{POWHEG-BOX-V2}}

\newcommand{\MINLO}{\noun{MiNLO}}
\newcommand{\LHE}{\noun{LHE}}

\newcommand{\HZJMINLO}{\noun{HZJ-MiNLO}}

\newcommand{\HZNNLO}{\noun{HZNNLO}}
\newcommand{\HZNNLOPS}{\noun{HZNNLOPS}}

\newcommand{\MCFM}{\texttt{MCFM-8.0}}
\newcommand{\xNNLOPS}{\scalebox{0.75}{\texttt{NNLOPS}}}
\newcommand{\xNNLO}{\scalebox{0.75}{\texttt{NNLO}}}
\newcommand{\xNLO}{\scalebox{0.5}{\text{NLO}}}
\newcommand{\xMINLO}{\scalebox{0.75}{\texttt{MiNLO}}}
\newcommand{\HZJ}{\scalebox{1.0}{\texttt{HZJ}}}

\newcommand{\HZNNLOLHE}{\noun{HZ-NNLOPS(LHEF)}}

\newcommand{\NNLOPS}{\noun{NNLOPS}}

\newcommand{\PYTHIA}[1]{\noun{Pythia{#1}}}

\newcommand{\HV}{\textrm{HV}}
\newcommand{\HW}{\textrm{HW}}
\newcommand{\HZ}{\textrm{HZ}}
\newcommand{\HVjet}{\textrm{HV+jet}}

\newcommand{\HZjet}{\textrm{HZ+jet}}
\newcommand{\Hbb}{{\textrm{H}}\!\to\!b\bar{b}}

\newcommand{\as}{\alpha_{\scriptscriptstyle \mathrm{s}}}

\newcommand{\mur}{\mu_{\scriptscriptstyle \mathrm{R}}}

\newcommand{\lm}{\ell^{-}}
\newcommand{\lp}{\ell^{+}}

\newcommand{\bb}{b\bar{b}}

\definecolor{darkgreen}{rgb}{0,0.6,0}
\definecolor{darkpurple}{rgb}{0,0.5,0.5}
\definecolor{darkblue}{rgb}{0,0,0.7}
\definecolor{darkred}{rgb}{0.5,0,0.0}
\definecolor{darkorange}{rgb}{0.8,0.4,0.0}
\definecolor{green}{rgb}{0.0,0.8,0.4}

\newcommand{\thetacs}{\theta^*}
\newcommand{\phics}{\phi^*}
\newcommand{\mhz}{M_{\sss \HZ}}
\newcommand{\yhz}{y_{\sss \HZ}}
\newcommand{\cosx}{\cos\!\alpha}

\newcommand{\PhiHZsimp}{\Phi_{\sss \HZ}}
\newcommand{\PhiHZtot}{\Phi_{\sss\rm{H}\ell\bar\ell}}
\newcommand{\PhiHZwhole}{\Phi_{\sss \ell\bar\ell b\bar{b}}}
\newcommand{\PhiHbb}{\Phi_{\sss H \to b\bar{b}}}
\newcommand{\ggHZ}{gg\!\to\!\rm{HZ}}

\newcommand{\mll}{M_{\sss \ell\bar\ell}}
\newcommand{\mbb}{M_{b\bar b}}

\newcommand{\gam}[1]{ \Gamma_{\sss b\bar b}^{(#1)} }

\newcommand{\dgam}[1]{ d\Gamma_{\sss b\bar b}^{(#1)} }
\newcommand{\dsig}[1]{ d\sigma^{(#1)} }
\newcommand{\BrHbb}{ {\rm{Br}}({\rm{H}}\!\rightarrow\! b\bar{b}) }
\newcommand\HH[1]{ H_{#1} }
\newcommand{\mh}{M_{\rm{H}}}
\newcommand{\mz}{M_{\rm{Z}}}
\newcommand\mzsq{M_{\rm{Z}}^{2}}
\newcommand\mhsq{p_{\rm{H}}^2}
\newcommand\gzsq{\Gamma_{\rm{Z}}^{2}}
\newcommand\qk[1]{ q_{#1}\!\!\cdot\! k}

\DeclarePairedDelimiter\floor{\lfloor}{\rfloor}

\usepackage[mathscr]{euscript}

\preprint{\\\\CERN-TH-2018-082\\ LAPTH-014/18\\ OUTP-17-18P}

\title{{NNLOPS accurate associated HZ production with NLO decay H$\to\! b\bar{b}$}}

\author[a]{William Astill,}
\author[a]{Wojciech Bizo\'n,}
\author[b,c]{Emanuele Re,}
\author[c]{Giulia Zanderighi\footnote{Rudolf Peierls Centre for Theoretical Physics,
  Clarendon Laboratory,
  Parks Road,
  Oxford, UK}}

\affiliation[a]{Rudolf Peierls Centre for Theoretical Physics,
  Clarendon Laboratory,
  Parks Road,
  Oxford, UK}

\affiliation[b]{LAPTh, CNRS, Universit\'e Savoie Mont Blanc, 74940 Annecy, France}

\affiliation[c]{Theoretical Physics Department, CERN, Geneva,
  Switzerland}

\emailAdd{william.astill@physics.ox.ac.uk}
\emailAdd{wojciech.bizon@physics.ox.ac.uk}
\emailAdd{emanuele.re@lapth.cnrs.fr}
\emailAdd{giulia.zanderighi@cern.ch}

\abstract{We present a next-to-next-to-leading order (NNLO) accurate
  description of associated \HZ{} production, followed by the Higgs
  boson decay into a pair of $b$-quarks treated at next-to-leading order
  (NLO), consistently matched to a parton shower (PS). The matching is
  achieved by performing reweighting of the \HZJMINLO{} events, using
  multi-dimensional distributions that are fully-differential in the
  \HZ{} Born kinematics, to the NNLO results obtained by using the
  \MCFM{} fixed-order calculation. Additionally we include the
  $\ggHZ{}$ contribution to the discussed process that appears at the
  $\mathcal{O}(\as^2)$.  We present phenomenological results obtained
  for 13 TeV hadronic collisions.}


\begin{document}
\maketitle \flushbottom

\section{Introduction}
\label{sec:intro}

After the discovery of the Higgs boson in
Run~I~\cite{Aad:2012tfa,Chatrchyan:2012xdj}, one of the main tasks of
the ongoing LHC~Run~II is to perform accurate measurements of Higgs
properties. In order to carry out this precision physics program, it is
important to study Higgs production in all the main production modes,
and compare measurements with theory predictions, for total
cross sections and differential distributions. An important goal which
is expected to be achievable with the Run~II full luminosity is to
establish solid statistical evidence for \HV{} associated
production~\cite{Aaboud:2017xsd,Sirunyan:2017elk}.

The past years have seen a remarkable progress in NNLO QCD
calculations, and, currently, all $2\to 2$ SM scattering processes are
known to this accuracy, see~\emph{e.g.}
ref.~\cite{Heinrich:2017una}. Thanks to this progress, the description
of colour singlet final states has reached a high level of
accuracy. This is particularly true for processes where, at
leading order (LO), there are no gluons in the initial state: in this case
higher-order corrections are typically moderate, and hence including
NNLO corrections leads to very stable results, with small perturbative
uncertainties. For \HZ{} production, the NNLO corrections have been
computed for the inclusive cross section~\cite{Brein:2003wg} as well
as for differential
distributions~\cite{Ferrera:2014lca,Campbell:2016jau,Caola:2017xuq}. Electroweak
corrections for this process are also known at NLO for inclusive
cross sections and differential
distributions~\cite{Ciccolini:2003jy,Denner:2011id}, and are
implemented in the public code {\tt
  HAWK}~\cite{Denner:2014cla}. Notably, NLO electroweak and QCD
corrections were simultaneously matched to a parton shower in
ref.~\cite{Granata:2017iod} for \HV{} and \HVjet{}, using, in the
latter case, the \MINLO{} method (to be described below).

Since associated \HV{} production has a small cross section, it is 
often considered in association with a Higgs boson decaying to a
$b$-quark pair, which is the largest Higgs decay mode. In this case
properties of the $b$-jets arising from the Higgs decay products are
used in experimental analysis to enhance the signal over SM
backgrounds, and they will also be important for extracting precise
information on the $b$-quark Yukawa coupling, especially in the Higgs
boosted regime. Because of this, including QCD corrections to the
$\Hbb{}$ decay is particularly important, especially since these
corrections are known to be large. The QCD NLO corrections to the
Higgs decay to massive $b$-quarks have been known for a long
time~\cite{Braaten:1980yq,Drees:1990dq,Sakai:1980fa,Janot:1989jf},
whereas more recently NNLO corrections were computed in
refs.~\cite{Anastasiou:2011qx,DelDuca:2015zqa} for massless
$b$-quarks. In the last few years the focus has moved towards a
combination of the aforementioned fully-differential NNLO computations
for $pp\to\HV{}$ with differential NLO and NNLO results for
$\Hbb{}$. The current state-of-the-art results are those obtained in
refs.~\cite{Ferrera:2017zex,Caola:2017xuq}, where the
fully-differential QCD NNLO computations for $pp\to\HV{}$ and $\Hbb{}$
(in the limit of massless $b$-quarks) have been combined together.

The precision of theory predictions is usually quantified in terms of
renormalisation and factorisation scale variation of the NNLO results.
It is however also known that all-order effects can be sizeable and
can give rise to effects that are outside the fixed-order scale
uncertainty band. For this reason, a lot of effort is put into
combining NNLO calculations with parton shower effects, thereby
obtaining so-called NNLOPS generators. Three methods have been
suggested recently to achieve this accuracy. The {\tt UNNLOPS} approach,
which has been used for Drell-Yan~\cite{Hoeche:2014aia} and Higgs
production~\cite{Hoche:2014dla}, is based on partitioning the phase
space into an unresolved and a one-jet region and a subsequent
matching to parton shower for the resolved one-jet region.
The {\tt Geneva} approach~\cite{Alioli:2013hqa} instead uses the
next-to-next-to-leading logarithmic (NNLL) accurate resummation for a
specific observable to essentially partition the phase space. This
method has been applied recently to Drell-Yan
production~\cite{Alioli:2013hqa}.
Finally, the \MINLO{} approach~\cite{Hamilton:2012np,Hamilton:2012rf}
relies on first using \MINLO{} to achieve an NLO merging of the
processes with the production of the colour-singlet state ($X$) and
the same processes with one additional jet ($X+1$ jet), and on
performing a reweighing of the \MINLO{} $X+1$ jet events to NNLO Born
distributions for $X$. This method has been applied recently to Higgs
production~\cite{Hamilton:2013fea,Hamilton:2015nsa}, Drell-Yan~\cite{Karlberg:2014qua} and \HW{} production~\cite{Astill:2016hpa}.

In this paper we consider the production of a Higgs boson in
association with a Z boson and consider the decay of the Higgs to
bottom quarks, the decay mode with the largest branching ratio, and
the decay of the Z boson to leptons. We build a Monte Carlo that is
NNLO accurate in production, preserves NLO accuracy in the decay and
includes parton shower effects.
We also include the NNLO $\ggHZ{}$ channel at leading order in
production, including LO corrections in decay and parton shower
effects. This subprocesses is added separately, and we assess its
numerical impact.
Our implementation uses and adapts the \POWHEGBOXRES{} code, which is
based on \POWHEGBOXVTWO{} but has a resonant-aware treatment of
internal resonances~\cite{Jezo:2015aia}, and hence it is suited to
treat NLO corrections to production and decay.

The paper is organised as follows. In Sec.~\ref{sec:method} we outline
the method used to obtain \NNLOPS{} accurate predictions including the
NLO treatment of the decay. In particular we explain how the latter is
included together with \MINLO{} within the \POWHEGBOXRES{}
framework. We detail how we parametrise the phase-space, and also
explain how we treat the ${\cal O}(\as^2)$ $\ggHZ{}$ contribution. In
Sec.~\ref{sec:practical} we give details about our practical
implementation, as well as about our interface to the parton
shower. In Sec.~\ref{sec:validation} we validate our results by
checking that we reproduce NNLO results for Born-like observables,
also for variables not used for the reweighting. In
Sec.~\ref{sec:pheno} we present distributions with fiducial cuts
inspired by the recent ATLAS analysis of
ref.~\cite{Aaboud:2017xsd}. Finally, we present our conclusions in
Sec.~\ref{sec:conclu}. A number of Appendices provide more details
about the treatment of the decay at NLO, the spectral decomposition
that we use to parametrise the phase space, the dependence of the
matrix element on the extra polar angle used in the phase-space
parametrisation, and the impact of $\ggHZ{}$ contribution.

\section{Outline of the method}
\label{sec:method}

In this work we consider the production of a Higgs boson in
association with a Z boson, followed by the Z boson decay into a pair
of leptons and the Higgs boson decay into pair of $b$-quarks
\begin{equation}
  \label{eq:process}
  pp \longrightarrow HZ
  \longrightarrow
  \left(b\bar{b}\right)
  \left(\ell^{+}\ell^{-}\right).
\end{equation}
The decay of the Z boson is treated exactly with all spin correlations
between initial and final-state fermions taken into account. The decay
of the Higgs boson is treated in the narrow-width approximation at
next-to-leading order in QCD.

In order to achieve \NNLOPS{} accuracy we follow the method of
reweighting Les Houches events (\LHE{}), produced by \MINLO{} improved
\HZJ{} generator (\HZJMINLO{}), with NNLO accurate fixed-order
predictions, differential in the Born phase space.  The procedure was
first proposed in~\cite{Hamilton:2012rf} and later implemented for
various colour-singlet production
processes~\cite{Hamilton:2013fea,Karlberg:2014qua,Hamilton:2015nsa,Astill:2016hpa}.
In its simplest implementation, the method consists in reweighting
\LHE{} event samples obtained with the \HZJMINLO{} generator using
multi-differential \HZNNLO{} distributions, with the factor
\begin{equation}
  \label{eq:rwgt-k-factor}
  \mathcal{W}(\PhiHZwhole{})
  =
  \frac{ d\sigma_{\xNNLO{}}(\PhiHZwhole) }{ d\sigma_{\xMINLO{}}(\PhiHZwhole) },
\end{equation}
where $\PhiHZwhole{}$ is the Born phase-space of
process~\eqref{eq:process}. The resulting event sample (which we refer
to by \HZNNLOLHE{}) is NNLO accurate: by construction the method
provides NNLO accuracy for all Born distributions and 1-jet
observables remain NLO accurate since the reweighting factor differs
from one by ${\cal O}(\alpha_s^2)$ corrections.
For the proof we refer the reader
to~\cite{Hamilton:2013fea,Karlberg:2014qua}. Furthermore, a subsequent
parton shower will not spoil the claimed accuracy provided that the
hardest real radiation for each event is generated by \POWHEG{}
itself.
This procedure was applied recently to \HW{} production in
ref.~\cite{Astill:2016hpa}, hence we refer the reader to that paper
for further details.
Instead, in this section, we first give a detailed description of the
treatment of the NLO $\Hbb{}$ decay, which is the new element of this
work (Sec.~\ref{sec:hbb-nlo-dec}). We then give some technical details
on how the reweighting to NNLO was achieved
(Sec.~\ref{sec:phase-space-decomp}
and~\ref{sec:reweighting-procedure}), and, finally, in
Sec.~\ref{sec:gghz-treatment} we discuss the inclusion of the
loop-induced $\ggHZ{}$ process, which is part of the NNLO corrections
to $pp\to\HZ{}$.

\subsection{\HZJMINLO{} with $\Hbb{}$  decay at NLO}
\label{sec:hbb-nlo-dec}

In this work we use the \MINLO{} prescription only for the production
part of the process, and match this to a ``resonance improved''
\POWHEG{} implementation of the NLO QCD calculation of the $\Hbb{}$
decay.
As described in the previous subsection, we treat the Higgs boson
decay in the narrow-width approximation (NWA).

We start by introducing the $\bar B$ function~\cite{Frixione:2007vw} that
we use in our \HZJ{} code, which reads schematically
\begin{equation}
  \begin{split}
    \bar B &=
    \alpha_s(q_t^2) \Delta_q^2(Q,q_t)
    \frac{{
        \BrHbb{} }}{\Gamma_{\xNLO}}
    \left[
      B_{\sss \rm HZJ} (1 - 2 \Delta_q^{(1)}(Q,q_t))
      \dgam{0} \right.
      \\ &
      + \left(V_{\sss \rm HZJ} +\int d
      \phi_r R_{\sss \rm HZJ}(\phi_r) \right) \dgam{0}
      + \left.  B_{\sss \rm HZJ} \left(\dgam{V}
      + \int d\phi_r \dgam{R}(\phi_r)\right) \right]\,,
\end{split}
\label{eq:minlo1} 
\end{equation}
where the Higgs propagator is left implicit; $\BrHbb{}$ is the best
prediction for the Standard Model $\Hbb{}$ branching ratio; and
$\dgam{0/V/R}$ are the Born, virtual, and real squared amplitudes for
the $\Hbb{}$ decay, differential in their kinematics.
$\Gamma_{\xNLO{}} = \gam{0} + \gam{1}$ denotes the NLO accurate
$\Hbb{}$ partial decay width.
With $\Delta_q(Q,q_t)$ we denote the \MINLO{} Sudakov form factor for
quark induced boson production (see ref.~\cite{Hamilton:2012rf} for
its precise definition) and $\Delta_q^{(1)}(Q,q_t)$ is its ${\cal
  O}(\alpha_s)$ expansion.
The hard scale in the Sudakov is set to $Q^2 = (p_Z+p_H)^2$ and $q_t$
is the transverse momentum of the $\HZ{}$ system, where the Higgs is
obtained from the sum of the momenta of its decay products ($b \bar b$
or $b \bar b g$).
In the formula above the additional $\alpha_s$ factor in the NLO
correction is contained implicitly in the $V$ and $R$ functions, as
well as in $\dgam{V}$ and $\dgam{R}$.
In the former two, following the \MINLO{} prescription, we set the
central renormalisation scale to $\mu_R = q_t$, whereas for the decay
we set the central scale to $\mur= \mh$ since this is the natural
scale for the decay and no \MINLO{} procedure is applied to it (in
App.~\ref{App:higgs-nlo-decay} we denote as $\mu_r$ the
renormalisation scale for the decay).

If we integrate Eq.~\eqref{eq:minlo1} over the phase space of all final-state light
partons we obtain
\begin{eqnarray}
 \label{eq:MiNLO-hz}
 \hspace*{-1.0cm}d\sigma_{\xMINLO}(\PhiHZwhole{}) &=& \BrHbb \!\cdot
 \!\left[ \left( \dsig{0}_{\rm \sss HZ} + \dsig{1}_{\rm \sss
     HZ}\right) \cdot\frac{ \dgam{0} +
     \dgam{1}}{\Gamma_{\xNLO}} \!+
   d\tilde{\sigma}^{(2)}_{\rm \sss HZ} \cdot \frac{\dgam{0}}{\Gamma_{\xNLO}} \right]+{\cal O}(\alpha_s^3),
\end{eqnarray}
where 
\begin{equation}
  \qquad
  \dgam{1} 
  =
  \dgam{V}
  + \int d\phi_r \dgam{R}(\phi_r)\,,\,
\end{equation}
and $d \sigma_{\rm \sss HZ}^{(i)}$ denotes the ${\cal O}(\alpha_s^i)$
correction to the \HZ{} production cross section.  The
$d\tilde\sigma^{(2)}$ denotes the ${\cal O}(\alpha_s^2)$ part of the
\HZJMINLO{} computation, which corresponds to double-real and
real-virtual parts of $\HZ{}$ production at NNLO.

We obtain NNLO prediction (\emph{without} the loop-induced $\ggHZ{}$
contribution, which is discussed in Sec.~\ref{sec:gghz-treatment}) for
the production combined with NLO corrections to the decay from
\MCFM{}, whose output is
\begin{eqnarray}
  \label{eq:NNLO-mcfm}
  d\sigma_{\xNNLO{}}(\PhiHZwhole{}) &=& \BrHbb{} \cdot \left[
    \dsig{0}_{\rm \sss HZ} \cdot \frac{\dgam{0} +
      \dgam{1}}{\Gamma_{\xNLO}} + (\dsig{1}_{\rm \sss
      HZ} + \,\dsig{2}_{\rm \sss HZ}) \cdot \frac{\dgam{0}}{\gam{0}} \right]\,.
\end{eqnarray}

It is easy to check that after integrating out the decay of the Higgs
boson in equation~\eqref{eq:NNLO-mcfm} one
recovers the fully inclusive NNLO result multiplied by the overall branching
ratio. One can also easily verify that
\begin{equation}
  \mathcal{W}(\PhiHZwhole{}) = \frac{
    d\sigma_{\xNNLO{}}(\PhiHZwhole{}) }{
    d\sigma_{\xMINLO}(\PhiHZwhole{}) } = 1 + \frac{\left(\sigma^{(2)}
    - \tilde{\sigma}^{(2)}\right) }{\sigma^{(0)}} +
  \mathcal{O}\left(\as^{3}\right)\,,
\end{equation}
which means that reweighting \HZJMINLO{} events with this factor does
not spoil the NLO accuracy of the event sample in the $\HZjet{}$
region, since the rescaling is equal to one up to $\mathcal{O}(\as^2)$
terms. In the following section we describe how we proceed to obtain
distributions differential in the Born phase space $\PhiHZwhole{}$.

\subsection{Phase-space parametrisation}
\label{sec:phase-space-decomp}
Our Born phase space contains four final-state particles, two leptons
($\lp, \lm$) and two bottom quarks ($b, \bar b$).
After neglecting an irrelevant azimuthal angle and upon inclusion of
the initial-state degrees of freedom we are left with 9 independent
dimensions. Furthermore we can factorise the Born phase-space as
follows:
\begin{equation}
   d\PhiHZwhole = d\PhiHZtot \times (2\pi)^3 dq^2\times d\PhiHbb,
\end{equation}
where $q^2$ is the virtuality of the Higgs boson, $\PhiHZtot$ is the
6-dimensional phase space for the production of an undecayed Higgs
boson with a pair of leptons from the decay of the associated Z boson,
and $\PhiHbb$ is the 2-dimensional phase space for the decay of a
Higgs boson into a pair of $b$-quarks.  By working in the NWA we
perform a reweighting only on the first part of the phase-space
$\PhiHZtot$, as the decay is already treated at the required accuracy
(NLO) in our implementation of \HZJMINLO{}.

A parametrisation of $\PhiHZtot{}$ can be defined in a number of
ways. After careful considerations we have chosen the invariant mass
and rapidity of the \HZ{} system ($\mhz$ and $\yhz$) to be the first
two variables. As a third variable we choose $\cosx$, where $\alpha$
is the polar angle of the Z boson with respect to the beam axis in the
frame where the \HZ{} system is at rest,~\emph{i.e.}
\begin{equation}
  \label{eq:cosx-def}
  \cosx = \frac{ \vec{p}\,'_{Z} \cdot \hat{z}' }{ |\vec{p}\,'_Z|\; |\hat{z}'| },
\end{equation}
where $'$ indicates that directions are expressed in the \HZ{}
rest-frame and the original $\hat{z}$ direction is along the beam
axis. Subsequently we choose the invariant mass of the Z boson,
$\mll{}$, and a convenient choice for the last two dimensions is to use
Collins-Soper angles ($\thetacs,\phics$) defined
in~\cite{Collins:1977iv}. In summary the full phase space parametrisation reads
\begin{equation}
  \label{eq:ps-coord}
  \PhiHZtot = \left\{ \mhz, \yhz, \cosx, \mll{}, \thetacs, \phics \right\}.
\end{equation}

Following the arguments of~\cite{Collins:1977iv} and the discussion in
Sec.~2 of~\cite{Astill:2016hpa} we
parametrise the ($\thetacs,\phics$)-dependence as follows:
\begin{eqnarray}
  \label{eq:diff-xs}
  \frac{d\sigma}{d\PhiHZtot} &=& \frac{d^6\sigma}{d\mhz{}\, d\yhz{}\,
    d\cosx\, d\cos \thetacs d\phics} \nonumber \\ &=& \frac{3}{16\pi}
  \left ( \frac{d \sigma}{d\PhiHZsimp}(1+\cos^2\thetacs) +
  \sum_{i=0}^{7} A_i(\PhiHZsimp ) f_i(\thetacs, \phics) \right)\,,
\end{eqnarray}
where in the second line we have used a short notation for the phase-space
without leptonic angles
\begin{equation}
  \label{eq:phase-space-3}
  \PhiHZsimp = \left\{ \mhz, \yhz, \cosx \right\}.
\end{equation}
The complete set of functions $f_{i}(\thetacs,\phics)$, together with
a procedure for extracting coefficients $A_{i}(\PhiHZsimp)$, is given
in equations (2.3-2.4) of ref.~\cite{Astill:2016hpa}. We note that for
practical purposes we will use only the first five $A_i$ coefficients
($A_0, \ldots A_4$) and, for simplicity, we neglect the remaining
three ($A_5,A_6,A_7$) since their impact on any distribution that we
examined is numerically negligible.

Finally, we can parametrise the dependence on the Z boson polar angle
$\alpha$ (see Eq.~\eqref{eq:cosx-def}) using a set of orthonormal
functions $g_{j}(\cosx)$. The definition of the basis elements $g_j$
is given in App.~\ref{App:GramSchmidt}. With this choice we have
\begin{eqnarray}
  \frac{d \sigma}{d\PhiHZsimp} = \sum_{j=0}^{N} c_j(\Phi) \,
  g_j\left(\cosx\right), \notag\\ A_{i}\left(\PhiHZsimp\right) =
  \sum_{j=0}^{N} a_{ij}(\Phi) \, g_j\left(\cosx\right),
  \label{eq:poly-decomp}
\end{eqnarray}
where $c_j$ and $a_{ij}$ are coefficients depending on $\Phi=\left\{
\mhz, \yhz \right\}$ and $N$ is the upper limit of the sum, which in
general can be inferred by analysing the matrix elements contributing
to the cross section.  We investigate the matrix elements in
App.~\ref{App:HadronicTensor} and find that, at most, polynomials of
5th-degree in the $\cosx$ and $\sin{\!\alpha}$ variables can
appear. Accordingly, we set $N=10$ in Eq.~\eqref{eq:poly-decomp}.

To summarise, our parametrisation of the fully differential cross
section as used in the reweighting procedure reads
\begin{equation}
  \label{eq:diff-reconstruction}
  \frac{d\sigma}{d\PhiHZtot} = \frac{3}{16\pi} \sum_{j=0}^{10}\left (
  c_j(\Phi)\,(1+\cos^2\thetacs) + \sum_{i=0}^{7} a_{ij}(\Phi)\,
  f_i(\thetacs, \phics) \right)g_{j}\left(\cosx\right)\,,
\end{equation}
where the functions $g_j(\cosx)$ are defined in Eq.~\eqref{eq:gdef}
and the coefficients $c_j(\Phi)$ and $a_{ij}(\Phi)$ can be obtained
from Eq.~\eqref{eq:poly-decomp} by exploiting the orthonormality of
the $g_j$ functions.

\subsection{Reweighting procedure}
\label{sec:reweighting-procedure}
The reweighting procedure leaves some degree of freedom for the
user. The simple rescaling with a factor, presented
in~\eqref{eq:rwgt-k-factor}, spreads the corrections uniformly over
the whole phase-space. However we know that regions where the
\HZ{} system is accompanied by hard QCD radiation is equally well
described by both predictions, \HZNNLO{} and \HZJMINLO{}. Hence, it is
desirable to limit the corrections to the phase space with no hard
jet.
To achieve this, we proceed along the lines of the prescription
presented in~\cite{Hamilton:2013fea}. We split the cross section into
two parts
\begin{equation}
  \label{eq:sigmaA-sigmaB}
  d\sigma_A = d\sigma\, h(p_t), \qquad d\sigma_B = d\sigma \left(1 - h(p_t) \right),
\end{equation}
where
\begin{equation}
  \label{eq:h-function}
  h(p_t) = \frac{ (\mh+\mz)^2 }{ (\mh+\mz)^2 + p_t^{\,2} }\,,
\end{equation}
and $p_t$ is the transverse momentum of the leading jet, computed here
using the $k_t$-algorithm with $R=0.4$. With such a choice
Eq.~\eqref{eq:rwgt-k-factor} takes form
\begin{eqnarray}
  \label{eq:W-rwgt}
  \mathcal{W}(\PhiHZtot{},p_t) &=& h(p_t)\,\frac{ \int
    d\sigma_{\xNNLO{}} \,\delta(\PhiHZtot - \PhiHZtot(\Phi)) - \int
    d\sigma_{\xMINLO{},B}\, \delta(\PhiHZtot - \PhiHZtot(\Phi)) }{
    \int d\sigma_{\xMINLO{},A}\, \delta(\PhiHZtot - \PhiHZtot(\Phi)) }
  \notag\\ & & + \left(1-h(p_t)\right).
\end{eqnarray}
This procedure smoothly turns off the corrections when moving to
regions of phase space with hard emissions whilst preserving the total
cross section,
\begin{equation}
  \left(\frac{d\sigma}{d\PhiHZtot}\right)_{\xNNLOPS{}}
  =
  \left(\frac{d\sigma}{d\PhiHZtot}\right)_{\xNNLO{}}.
\end{equation}
It is worth noting that choosing the transverse momentum $p_t$ in
Eq.~\eqref{eq:h-function} as the transverse momentum of hardest jet is
not equivalent to choosing the transverse momentum of the
colour-singlet, the difference being due to configurations where the
colour-singlet has small transverse momentum and it is accompanied by
hard QCD emissions whose transverse momenta counterbalance each other.
The latter configurations dominate in the region where $p_{t,\HZ{}} \sim
0$.

In the next section we will explain how we have included in our
simulation the loop-mediated $\ggHZ{}$ contribution, which was not
included in ${d\sigma}_{\xNNLO{}}$ (and hence not even in the
reweighting factor defined in Eq.~\eqref{eq:W-rwgt}), whilst being
formally $\mathcal{O}(\as^2)$.

\subsection{Treatment of the $\ggHZ{}$ contribution}
\label{sec:gghz-treatment}
The $\mathcal{O}(\as^2)$ contributions of the form $\ggHZ{}$, that
appear in the process of Higgs boson production in association with
a Z boson, originate from the squared 1-loop diagrams
shown for example in Fig.~5 of ref.~\cite{Campbell:2016jau}.
There are two classes of contributions: box diagrams involving a Yukawa
coupling of the Higgs boson, and triangle diagrams where the Higgs
boson is radiated from the Z boson which couples to the fermion
loop.
Both contributions vanish trivially when massless quarks of the first
and second generation run in the loop. 
These $\ggHZ{}$ contributions are known to constitute a significant
part of the cross
section~\cite{Hespel:2015zea,Goncalves:2015mfa,Ferrera:2014lca,Campbell:2016jau,Ferrera:2017zex},
especially when the invariant mass of the produced \HZ{} system is
larger than twice the top-quark mass. Being loop-induced, so far only
approximate NLO corrections are known for this
channel~\cite{Altenkamp:2012sx,Harlander:2014wda}. On the other hand,
their loop origin makes these terms particularly sensitive to New
Physics~\cite{Englert:2013vua,Harlander:2013mla,Harlander:2018yns}.

In our reweighting procedure we do not include this contribution, but
we will treat it independently using a separate event sample produced
using a leading order implementation of loop-induced $\ggHZ{}$ process
implemented in \POWHEG{}~\cite{Luisoni:2013kna}, which includes top
and bottom quarks in the loop.\footnote{The code can be obtained from
  \texttt{svn://powhegbox.mib.infn.it/trunk/User-Processes-V2/ggHZ}.}
We note that the $\ggHZ{}$ contribution can be treated separately
since it is finite.  Furthermore, parton shower radiation from
gluon-initiated hard processes is typically different from processes
also involving initial-state quarks. From that point of view, it is
important not to include the $\ggHZ{}$ contribution through a simple
reweighting.
Further discussion of the $\ggHZ{}$ channel is presented in
App.~\ref{App:gghz}.
For this contribution we do not include any radiative correction to
the the $\Hbb{}$ decay, hence the radiation from the decay is
fully taken care of by \PYTHIA{8}. Higher-order NLO corrections to this
decay could also be included with relatively little effort.

\section{Practical implementation}
\label{sec:practical}

In this section we first discuss the codes used to obtain our
predictions, as well as the relevant settings and the parameters. We
also describe subtleties related to the interface to \PYTHIA{8} when
radiation from resonances is taken into account. 

\subsection{Codes and settings}
\label{sec:setting}

In order to obtain an ensemble of NNLOPS accurate Les Houches events
for the process in Eq.~\eqref{eq:process} we need fully differential
predictions from an NNLO fixed-order calculation, and an NLO accurate
event-sample for \HZJ{} production improved with the \MINLO{}
prescription.

For the NNLO fixed-order prediction we use the \MCFM{}
calculation~\cite{Campbell:2016jau}. In order to obtain both the NNLO
accuracy for the production of the \HZ{} resonance as well as the NLO
accuracy of hadronic decay of Higgs boson, $\Hbb{}$, we
perform two separate runs of the program with \texttt{nproc=101} at
the \texttt{'nnlo'} order (for the first) and \texttt{nproc=1010} at
the \texttt{'nlo'} order (for the latter). The prediction presented in
Eq.~\eqref{eq:NNLO-mcfm} is simply obtained by adding results of the two runs.
As it was pointed out in Sec.~\ref{sec:hbb-nlo-dec} and
Sec.~\ref{sec:gghz-treatment}, we do not include $\ggHZ{}$
contributions in the reweighting procedure. To remove them we have
modified part of the \MCFM{} code, which computes double-virtual
corrections. We will include this contribution in our phenomenological
analysis in Sec.~\ref{sec:pheno}, as stated clearly in the appropriate
places.

The initial sample of Les Houches events is generated using the
\HZJMINLO{} package, originally in \POWHEGBOXVTWO{}, which we adapted
to run in \POWHEGBOXRES{}~\cite{Jezo:2015aia}.
We have also extended the original package to include NLO corrections
of the Higgs boson decay into pair of $b$-quarks, as discussed in
Sec.~\ref{sec:hbb-nlo-dec}. The relevant matrix elements have been
reported in App.~\ref{App:higgs-nlo-decay}.
Despite the fact that there is no interference between production and
decay, in order to treat the radiation from the resonance we have made
use of the \POWHEGBOXRES{} framework~\cite{Jezo:2015aia}.

In our work we use \texttt{PDF4LHC15\_nnlo\_mc} parton distribution
functions~\cite{Ball:2014uwa,Harland-Lang:2014zoa,Dulat:2015mca,Carrazza:2015hva}. We
set $\mh = 125.0$ GeV, $\Gamma_H = 4.088$ MeV, $\mz=91.1876$ GeV, and
$\Gamma_Z = 2.4952$ GeV. Moreover $G_F = 1.16639\cdot 10^{-5}
\rm{GeV}^{-2}$, $\sin^2\theta_{W} = 0.2223$,
$\alpha_{\scalebox{0.5}{\rm{EM}}}(\mz) = 128.89$, and $\BrHbb{} =
0.5824$.

For the contributions where the Higgs boson is radiated from a
heavy-quark loop we use pole mass of the heavy quark as it is usually
done in publicly available codes~\cite{Campbell:2016jau}. In
particular we set pole mass of the bottom quark to $m_b = 4.92$ GeV
and pole mass of the top quark to $m_t = 173.2$ GeV. Moreover, for the
bottom Yukawa coupling in $\Hbb{}$ decay we use its $\MSbar$ running
mass evaluated at scale $\mh$. The running masses are computed from
the pole masses using an ${\cal O}(\alpha_s^2)$
conversion~\cite{Chetyrkin:2000yt} and the numerical value of the
$\MSbar$ mass for the bottom quark, obtained by running the strong
coupling using \LHAPDF{}, is $m_b(\mh) = 3.16$ GeV.

The NNLO fixed-order prediction is obtained using fixed
renormalisation and factorisation scale equal to sum of the Higgs
boson and the Z boson mass, $\mu = \mh+\mz$. The scale choice in
\HZJMINLO{} case for the production is fixed by \MINLO{}
procedure~\cite{Hamilton:2012np}, while for the decay the central
scale choice is $\mh$, as explained in Sec.~\ref{sec:hbb-nlo-dec}.
We estimate the theoretical uncertainty using 7 point scale variation
for both fixed-order NNLO results as well as \MINLO{} predictions. The
scale variation in production and decay are always correlated (this
includes the decay renormalisation scale,~\emph{i.e.} the scale at which we
evaluate the $\MSbar$ $b$-quark mass).
We perform correlated variations in \MCFM{} and
\HZJMINLO{}, that is, our final uncertainty is an envelop of 7
scale combinations,~\emph{i.e.} for a given $(K_r,K_f)$ choice in \HZJMINLO{}
we use the same choice in \MCFM{} and as usual we consider variations
of the central scale by a factor two up and down, restricted to $1/2
\le K_r/K_f < 2$.
The $\ggHZ{}$ contribution is then added with the same $(K_r,K_f)$
choice.

When interfacing our fixed-order predictions to a parton shower we use
\PYTHIA{8}~\cite{Sjostrand:2014zea}, as detailed more precisely in the
next subsection. Unless stated otherwise, predictions are shown at
parton level, with no multi-parton interactions.

\subsection{Interface to Parton Shower}
\label{sec:intshower}
In order to combine our results with a parton shower we follow an
approach similar to the one first introduced for the NLO $t\bar t$
production treatment in \cite{Campbell:2014kua}, that allows for a
generation of radiation also from resonances. In our simulation we set
the flag \texttt{allrad} to 0, which means that the NLO \POWHEG{}
emission is generated at most from one singular region, associated
either with the production stage or with the radiation from a
resonance and we do not consider radiation from multiple regions.
This is the standard \POWHEG{} procedure to generate the hardest
radiation. In this configuration \POWHEG{} uses the usual highest bid
mechanism to choose the origin of the emission.

For the parton shower we use \PYTHIA{8}.  A requirement for the
matching to the parton shower to work properly is that the hardest
radiation should be the one generated by \POWHEG{}. This is usually
achieved by setting a value of \texttt{scalup} in every event, which
sets the veto scale for the parton shower. One subtlety is however
that the definition of the hardness of the radiation from the decay in
\POWHEG{} and \PYTHIA{8} differ. As a consequence, after an event is
showered, we recompute the hardness of the first emission generated by
\PYTHIA{8} using the \POWHEG{} formula and accept the showered event
only if this hardness is lower than the \texttt{scalup} value of the
given event. If this is not the case, we shower the event again until
the new showered event meets the required condition. Details of the
hardness definition used in \POWHEG{} and \PYTHIA{8} are given in
App.~A of ref.~\cite{Campbell:2014kua}. We have checked that our
procedure to veto radiation gives results that are fully compatible
with those obtained by the procedure encoded in the
\POWHEGHOOK{}-class provided by \PYTHIA{8}~\cite{Sjostrand:2014zea}. 

\section{Validation}
\label{sec:validation}

In the following section we present the validation of our results.  We
carefully compare distributions prepared from reweighted Les Houches
events (\HZNNLOLHE{}) with the ones obtained using fixed-order NNLO
code (\MCFM{}). We remind the reader that, in the reweighting
procedure, we don't take into account one-loop squared contributions
arising from $\ggHZ{}$ channel, as specified and motivated in
Sec.~\ref{sec:reweighting-procedure} and~\ref{sec:gghz-treatment}
respectively.  Therefore all the plots of this section do not contain
the $\ggHZ{}$ channel, which instead will be included in
Sec.~\ref{sec:pheno}.

In the plots of this section, the blue, green and red markers
represent results from \HZJMINLO{} Les Houches events, the fixed-order
calculation obtained with \MCFM{}, and the reweighted \HZNNLOLHE{} event
sample, respectively. The uncertainty band represents the usual scale
variation uncertainty, as described in detail in the previous section.

\begin{figure}
  \centering
  \includegraphics[page=1,width=205pt]{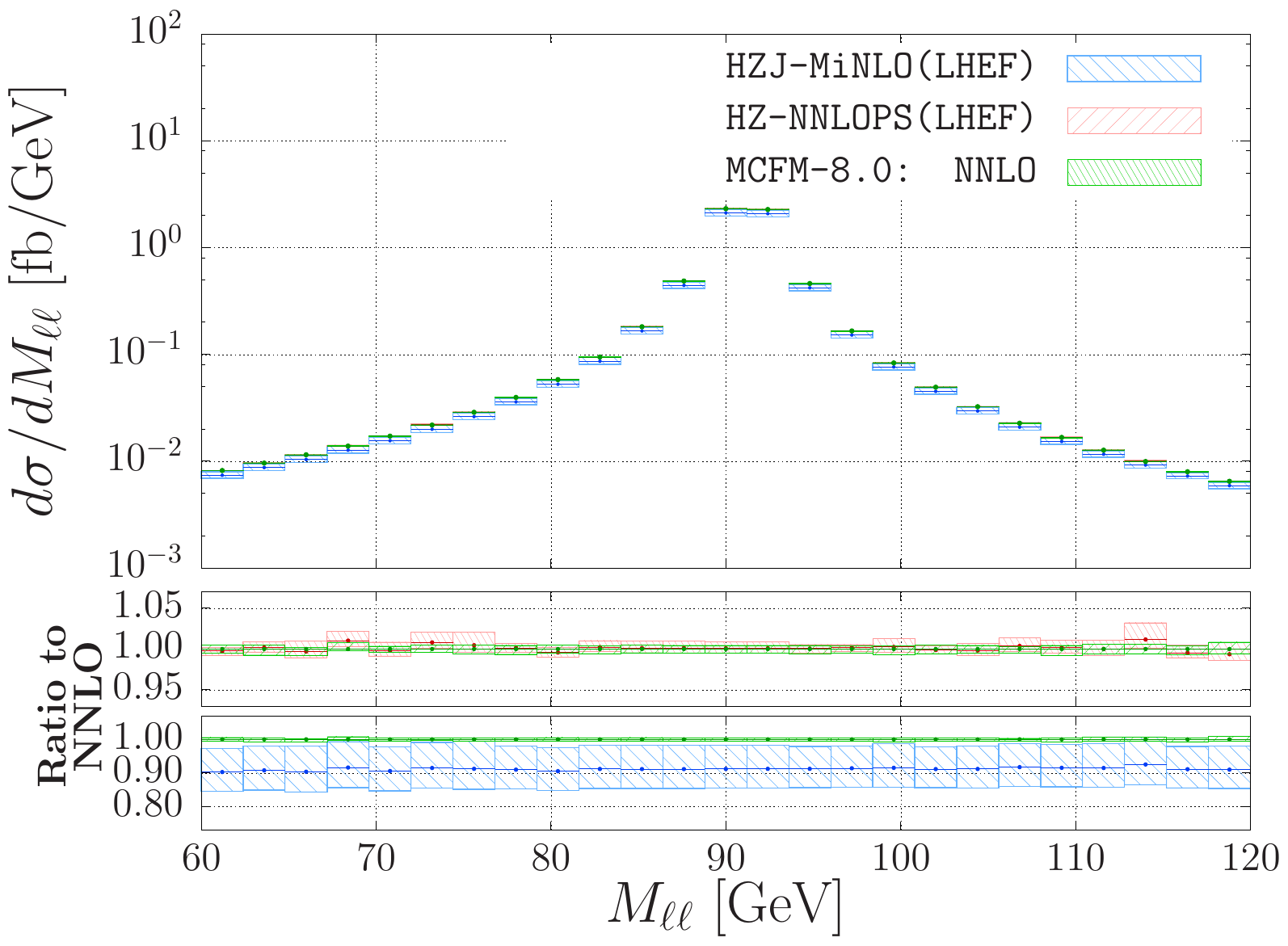}
  \hspace*{0.25cm}
  \includegraphics[page=1,width=205pt]{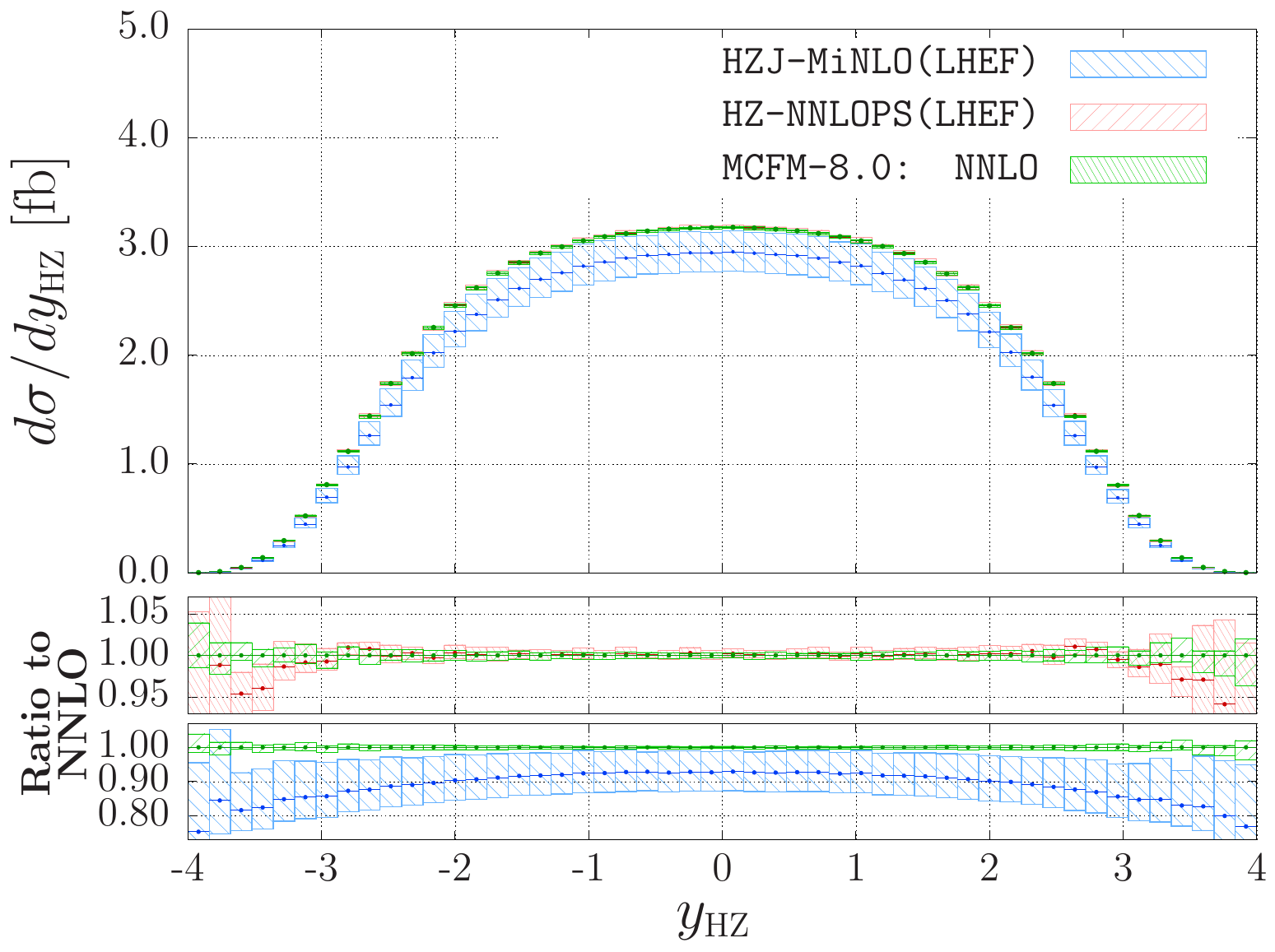}
  \caption{The differential distributions of the invariant mass of
    final-state leptons $\mll$ (left panel) and the distribution of
    the rapidity of the \HZ{} system $\yhz$ (right panel). The one-loop squared terms
    from the $\ggHZ{}$ channel have not been included.}
  \label{fig:mll-yHZ-plots}
\end{figure}

\begin{figure}
  \centering
  \includegraphics[page=1,width=205pt]{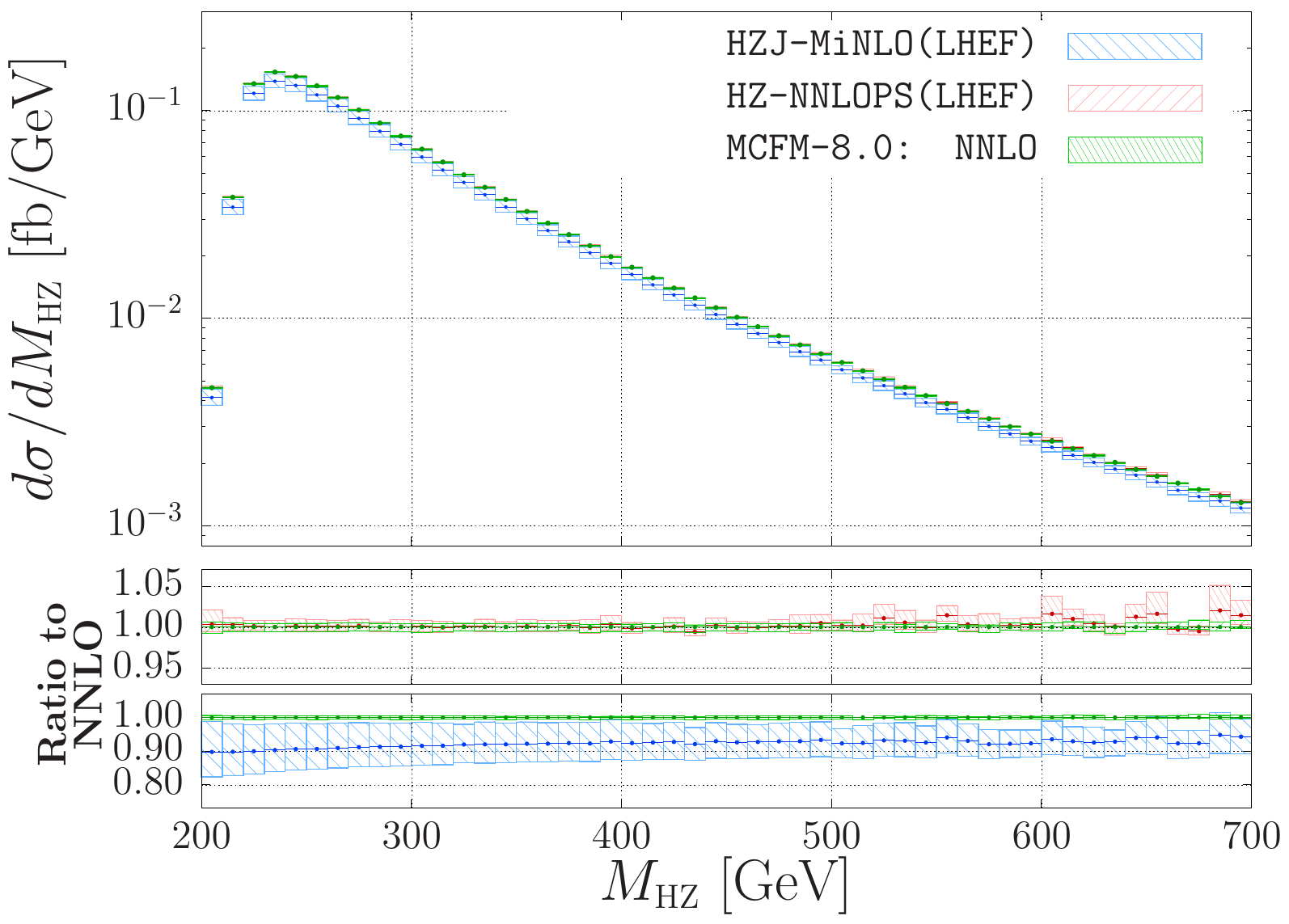}
  \hspace*{0.25cm}
  \includegraphics[page=1,width=205pt]{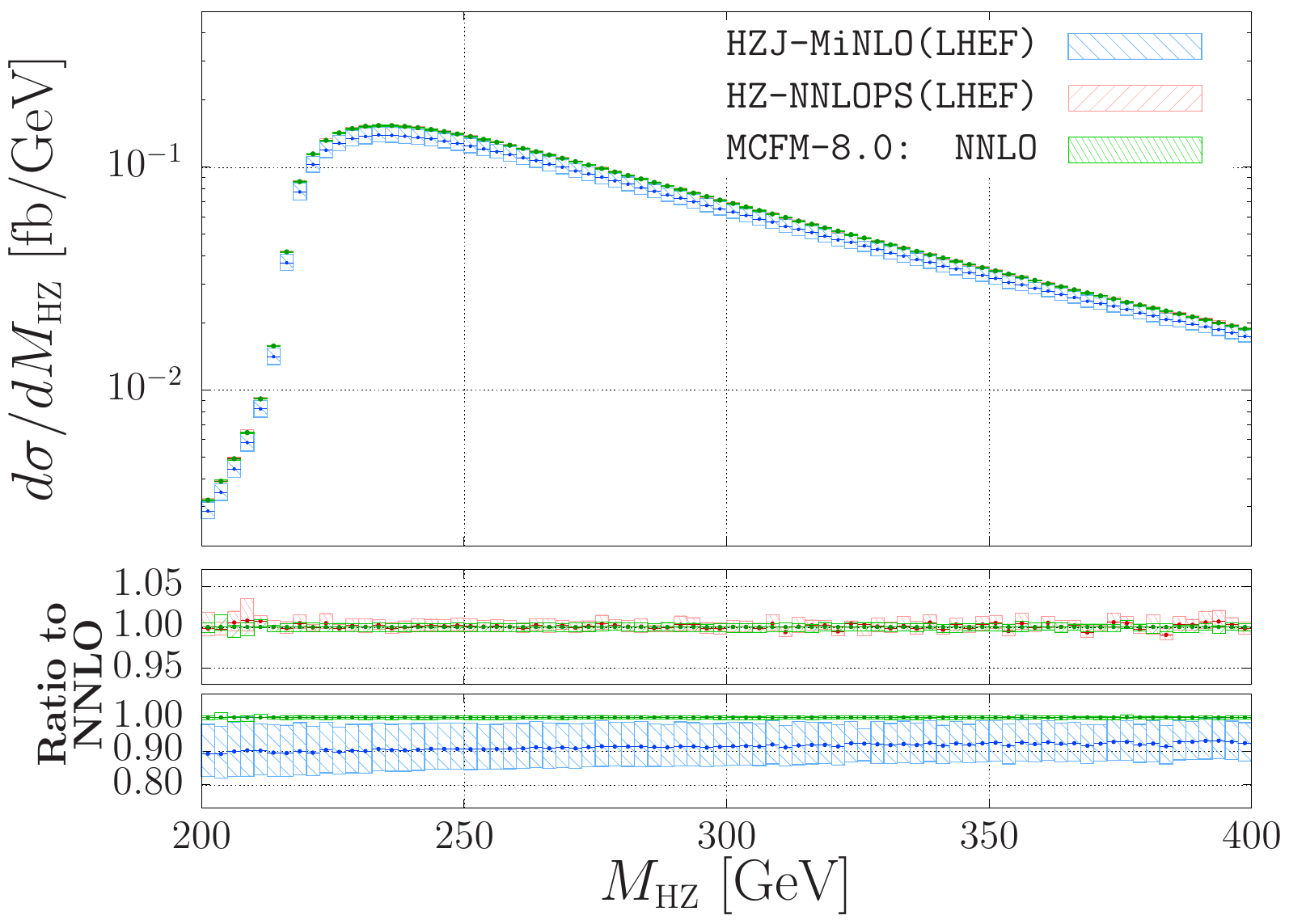}
  \caption{The differential distributions of the invariant mass of the
    \HZ{} system $\mhz$ in two different mass regions. The one-loop
    squared terms from $\ggHZ{}$ channel have not been included.}
  \label{fig:mHZ-plots}
\end{figure}

The first pair of plots that we want to present is the distribution of
the invariant mass of final-state leptons $\mll$ and the distribution
of the rapidity of the \HZ{} system $\yhz$, which are shown in the
left and right panel of Fig.~\ref{fig:mll-yHZ-plots}, respectively.
We start by noting that the ratio of \MCFM{} to \HZJMINLO{}, bottom
panel of Fig.~\ref{fig:mll-yHZ-plots}, is constant, which is along the
lines of our assumption that the reweighting factor
$\mathcal{W}(\PhiHZwhole{})$ should be constant along the $\mll$
direction. We do not repeat the thorough procedure of validation, which
was included in our previous work~\cite{Astill:2016hpa}.
The distribution of $\yhz$, right panel of
Fig.~\ref{fig:mll-yHZ-plots}, is again properly reproduced by our
calculation across the whole spectrum. We take note of the fact that
the discrepancies at the edges of the distribution are in the regions
of phase-space which are poorly populated. More precisely, having used
distributions with varying bin-size for the reweighting, all events
with $\yhz \lesssim -3$ (or $\yhz \gtrsim +3$) fall into the first (or
the last) bin of the differential reweighting factor
$\mathcal{W}(\PhiHZwhole{})$.
The description of the forward rapidity region can be improved by 
increasing the statistics of the multi-differential distributions and
by including more bins at large rapidity.
In Fig.~\ref{fig:mHZ-plots}, we present the differential distribution
of $\mhz$ in two different mass regions.
For this distribution the
difference between \HZJMINLO{} and the NNLO is small and flat over the
whole range. After reweighting, we find perfect agreement between NNLO
and \HZNNLOLHE{} results.

\begin{figure}
  \centering
  \includegraphics[page=1,width=205pt]{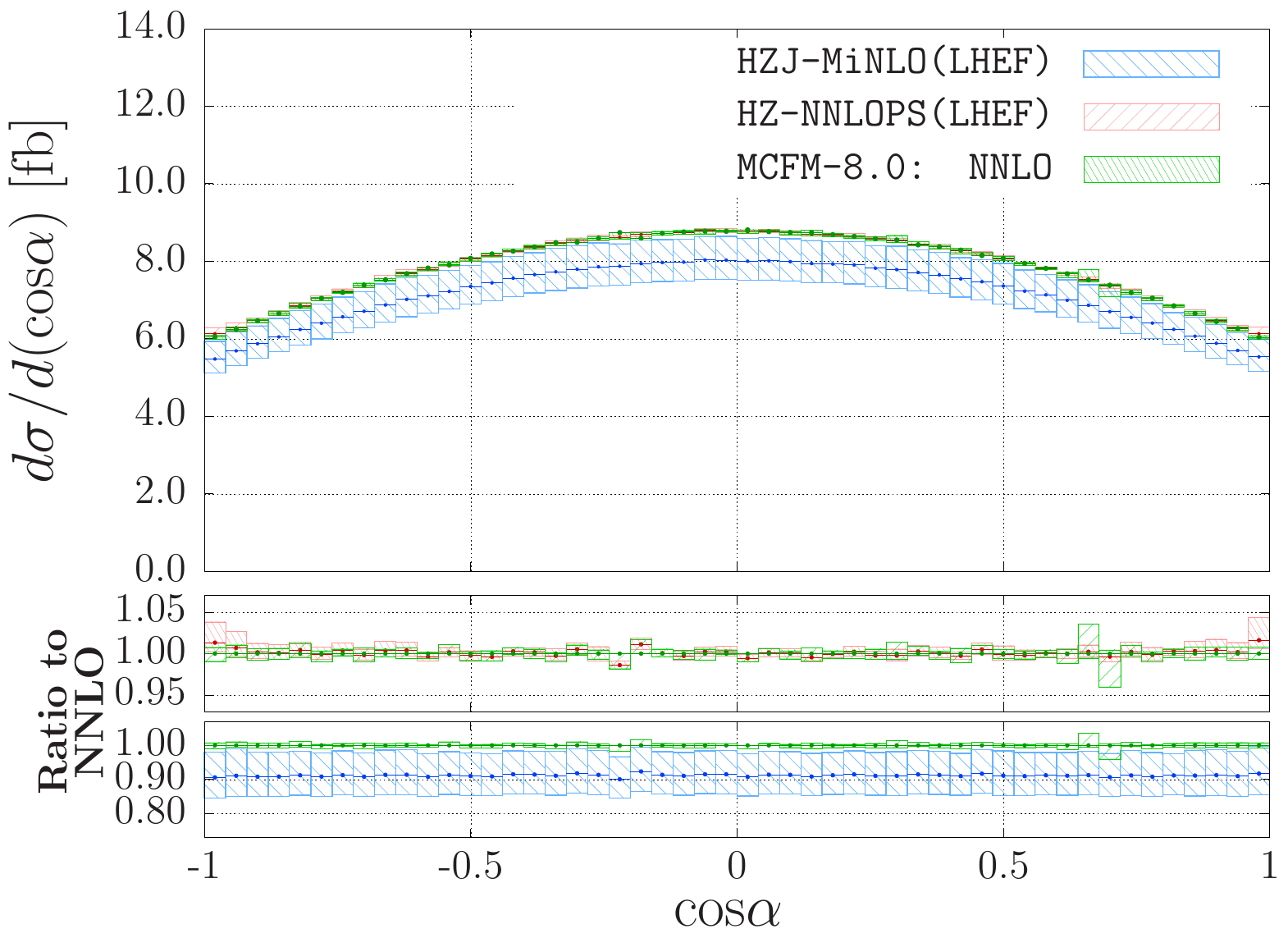}
  \hspace*{0.25cm}
  \includegraphics[page=1,width=205pt]{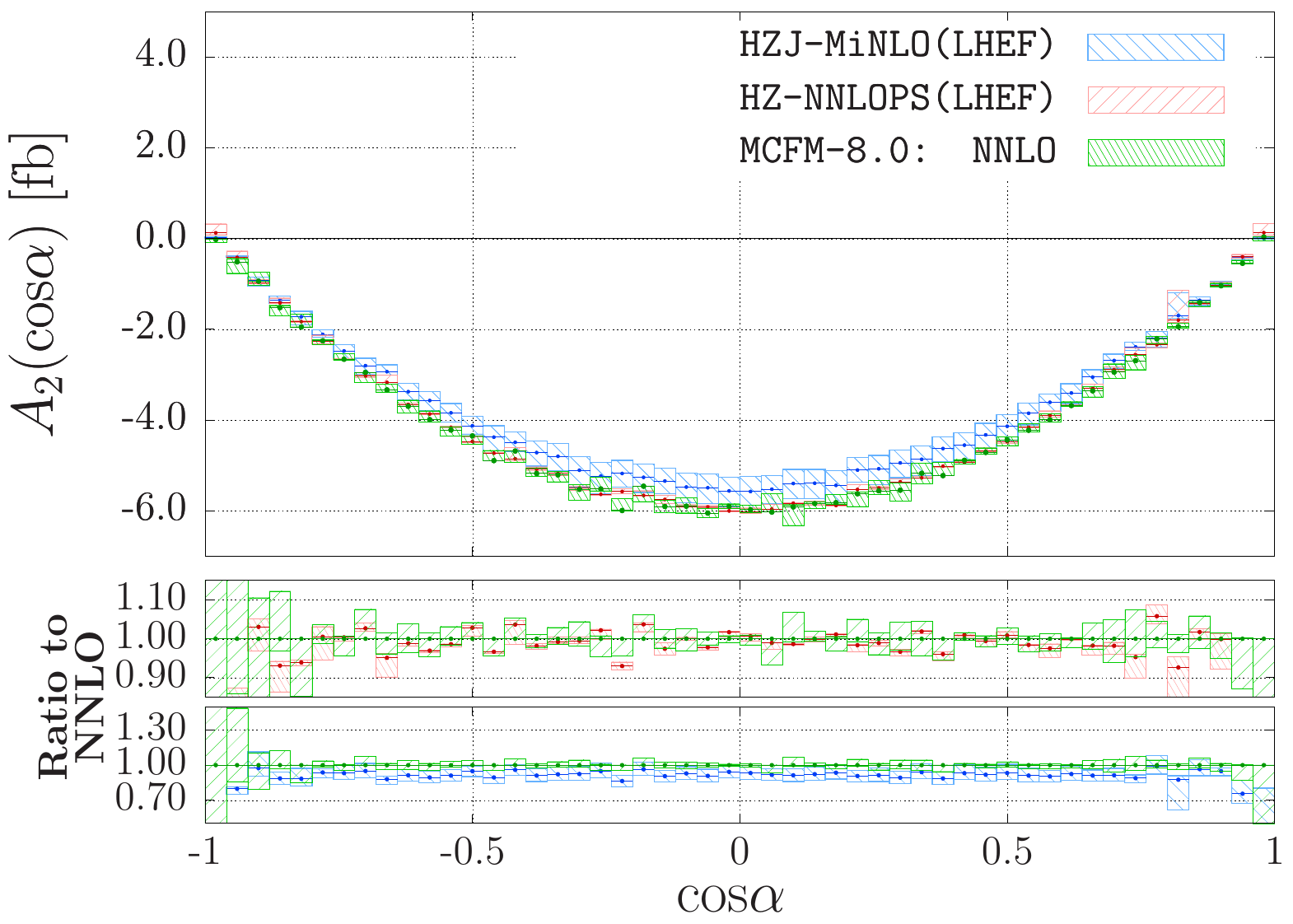}
  \caption{The differential distributions of the Z boson polar angle
    with respect to the beam axis, defined in Eq.~\eqref{eq:cosx-def}:
    differential cross section as a function of $\cosx$ (left panel)
    and the dependence of coefficient $A_2(\cosx)$, see
    Eq.~\eqref{eq:diff-xs}. The one-loop squared terms from the
    $\ggHZ{}$ channel have not been included.}
  \label{fig:cosx-plots}
\end{figure}

\begin{figure}
  \centering
  \includegraphics[page=1,width=205pt]{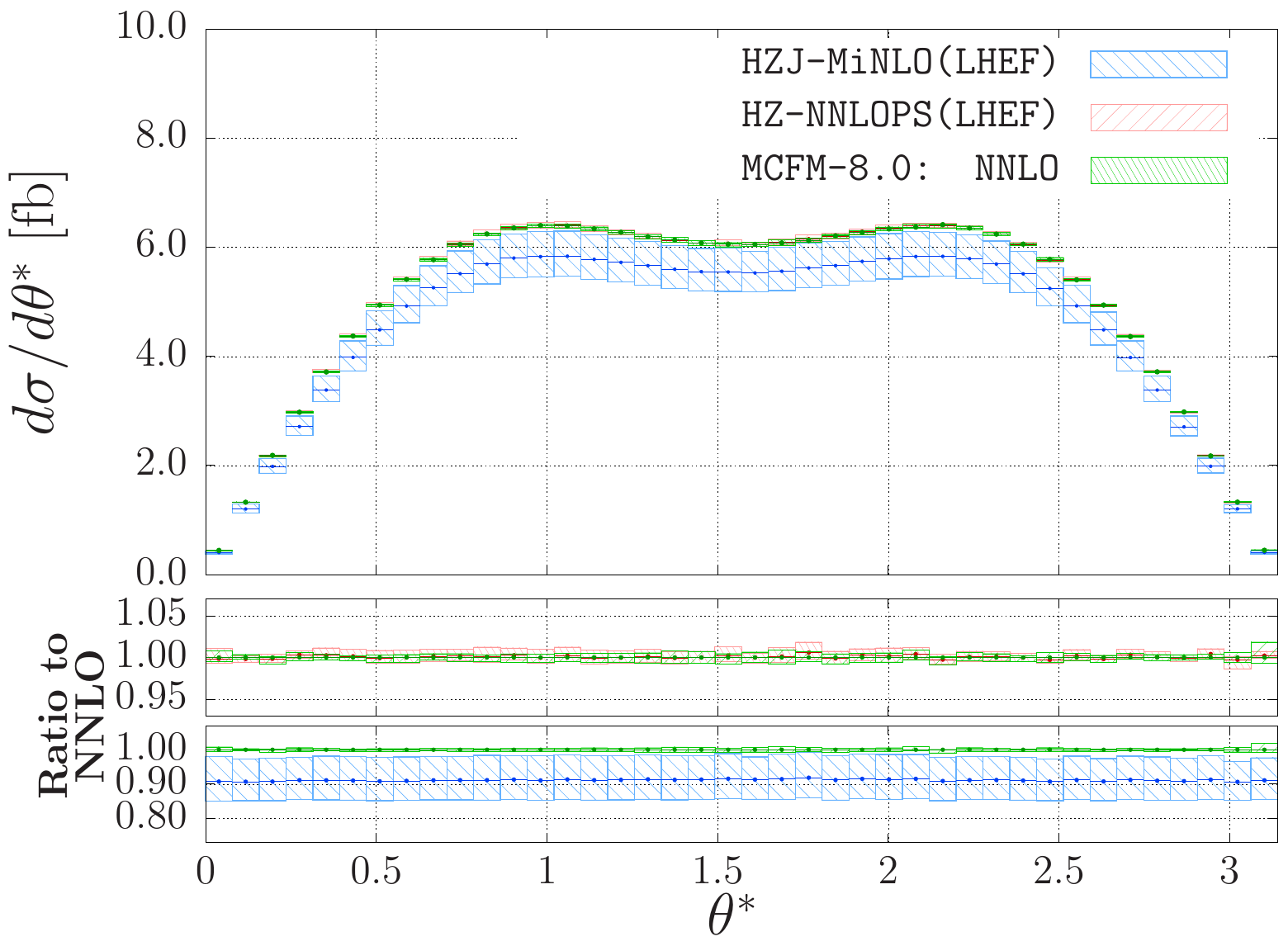}
  \hspace*{0.25cm}
  \includegraphics[page=1,width=205pt]{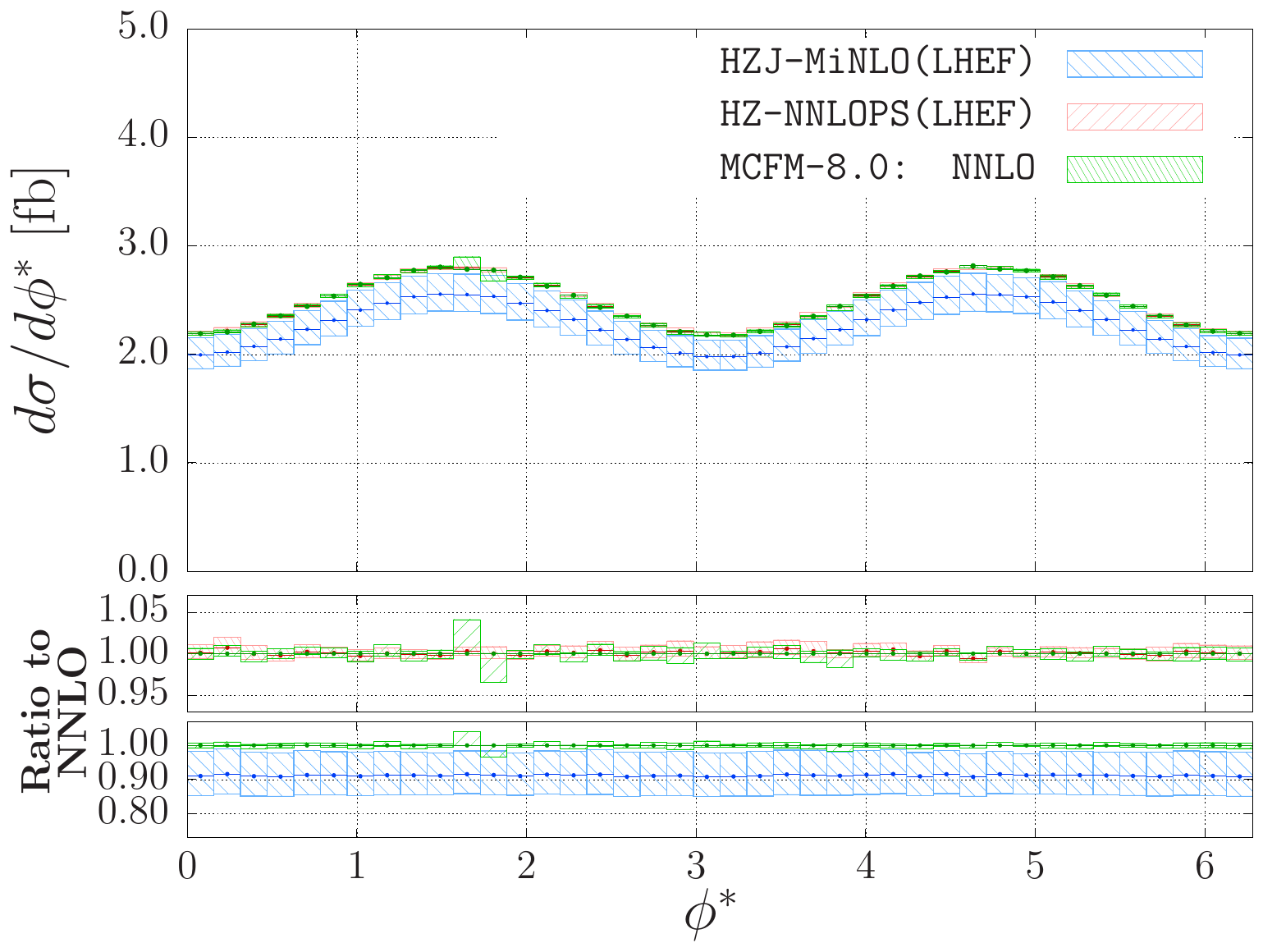}
  \caption{The differential distributions of Collins-Soper angles:
    $\thetacs$ (left) and $\phics$ (right). The one-loop squared terms
    from the $\ggHZ{}$ channel have not been included.}
  \label{fig:cs-angles}
\end{figure}

As the next step, we look closely at the differential distributions of
the angular variables: $\cosx$ and Collins-Soper angles. The
distribution of $\cosx$ is presented in Fig.~\ref{fig:cosx-plots}. We
recollect that this dependence was not just recorded as a histogram,
but rather parametrised in terms of spectral modes,
Eq.~\eqref{eq:poly-decomp}. This has improved the stability of the
distribution, and as a consequence of the reweighting factor, which is
very useful when working with samples of limited statistics.
Further we check the quality of the reconstruction of the
Collins-Soper angles. We present the relevant distributions in
Fig.~\ref{fig:cs-angles}.
In summary, we find that for all variables used for the reweighting,
the NNLO and \HZNNLOLHE{} predictions agree within their statistical
fluctuations.

\begin{figure}
  \centering
  \includegraphics[page=1,width=205pt]{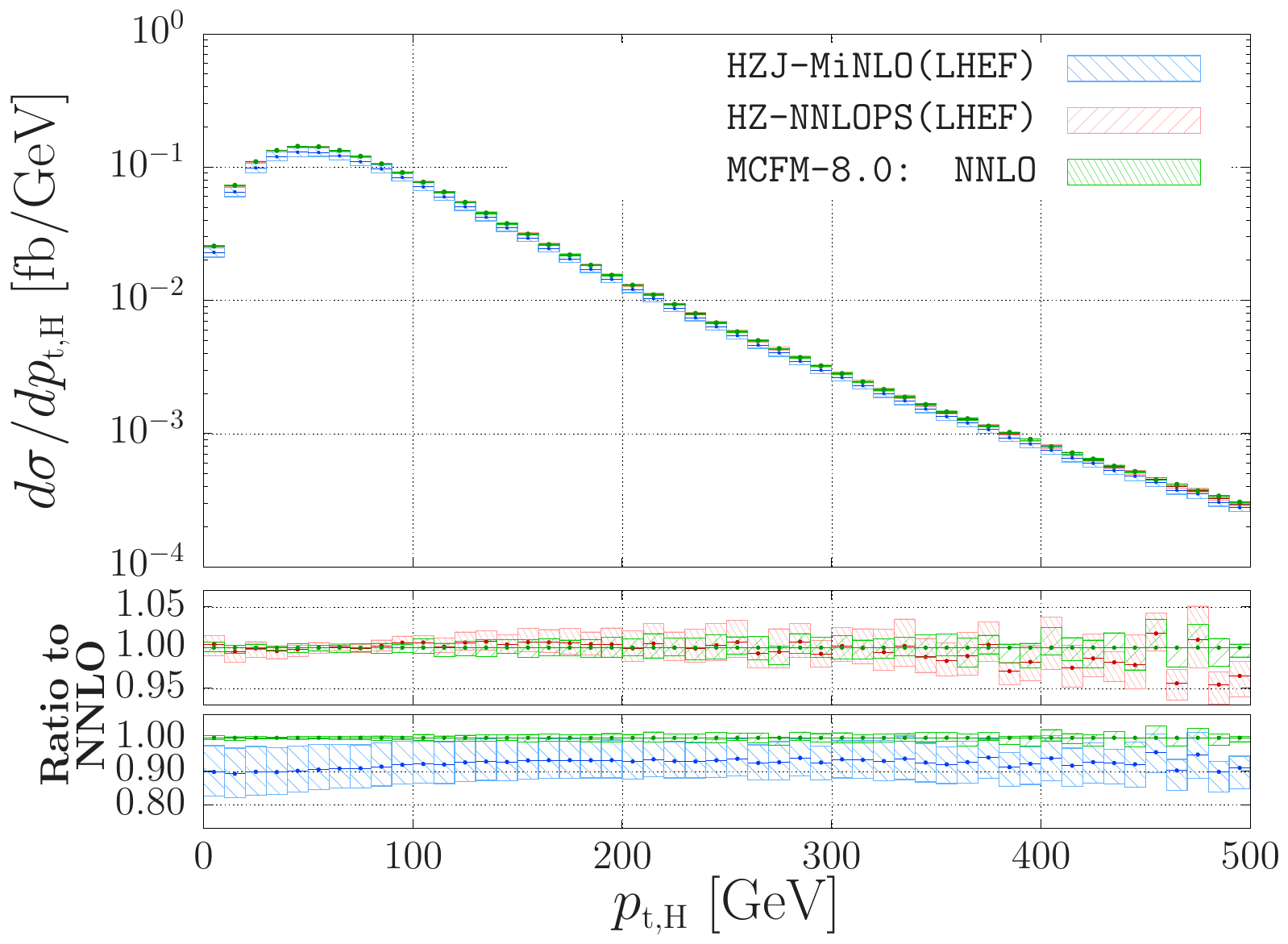}
  \hspace*{0.25cm}
  \includegraphics[page=1,width=205pt]{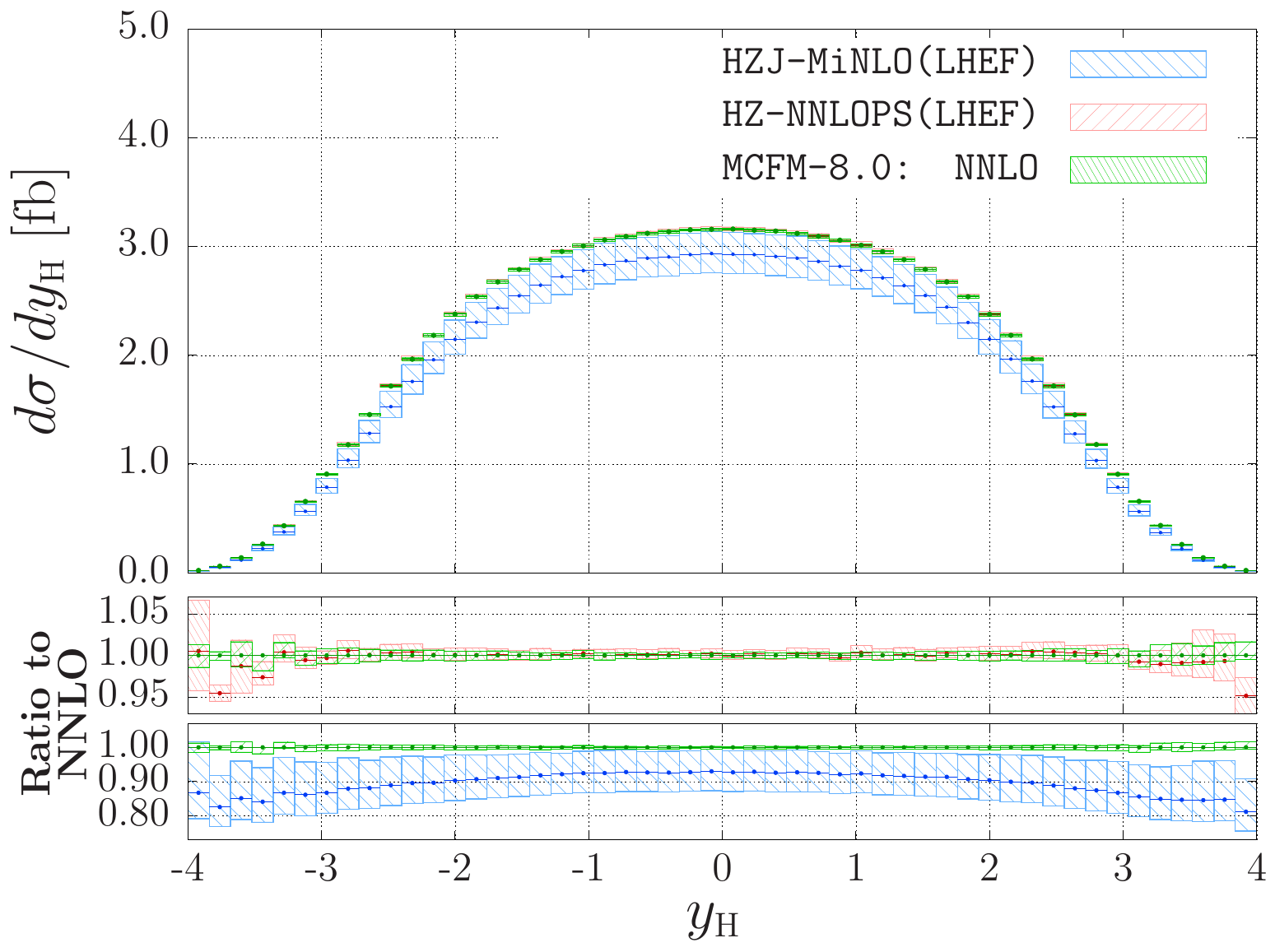}
  \caption{The differential distributions of the transverse momentum
    (left panel) and the rapidity (right panel) of the Higgs
    boson. The one-loop squared terms
    from the $\ggHZ{}$ channel have not been included.}
  \label{fig:higgs-plots}
\end{figure}

To complete the validation, we also need to examine Born-like
observables that were not used in the reweighting procedure. As such,
we chose to look at transverse momentum and rapidity of the Higgs
boson (see Fig.~\ref{fig:higgs-plots}). We again confirm that both the
central values and scale variation bands are properly reconstructed
within statistical fluctuations, which increase at high transverse
momentum ($p_{t,\rm{H}} \gtrsim 400$ GeV) and large rapidity ($|y_{\rm
  H}| \gtrsim 3$).
The agreement between NNLO and \HZNNLOLHE{} in these corners of phase space
could be improved further by increasing the statistics of the
reweighting factor and decreasing the bin-sizes in this region.

Finally we turn to the discussion of the distribution of the
transverse momentum of \HZ{} system, an observable which is singular
at Born level but receives corrections due to QCD radiation at
higher-orders in perturbation theory.
We compare results obtained using two different reweighting
prescriptions: the one described in
Sec.~\ref{sec:reweighting-procedure}, presented in the left plot of
Fig.~\ref{fig:pthz-plot}, and a setup where we set the function
$h(p_t)\equiv 1$ in Eqs.\eqref{eq:sigmaA-sigmaB}-\eqref{eq:W-rwgt},
shown in the right hand side of Fig.~\ref{fig:pthz-plot}.
As expected, we observe that the \HZJMINLO{} and \HZNNLOLHE{}
predictions feature a Sudakov damping at low transverse momentum,
while the NNLO prediction diverges in this region.
Furthermore, we observe that for the ${h(p_t)=1}$ case, the
\HZNNLOLHE{} results are uniformly shifted with respect to the
original event sample \HZJMINLO{}, as the reweighting factor
$\mathcal{W}(\PhiHZtot{})$ does not take into account any QCD
radiation. Instead, when the reweighting factor depends on the
transverse momentum of the leading jet, \HZNNLOLHE{} approaches the
\HZJMINLO{} curve at high-$p_{\rm t,HZ}$ values, as the effects of the
reweighting are concentrated in the region of phase-space close to the
Born kinematics, the natural habitat of $\mathcal{O}(\as^2)$ virtual
corrections.
In this case, the \HZNNLOLHE{} prediction at high transverse momentum
agrees with the \HZJMINLO{} prediction, rather than with the pure NNLO
result. We note that in this region, all predictions are only NLO
accurate and that the former has a dynamical scale, dictated by the
\MINLO{} prescription, while the NNLO uses a fixed renormalisation and
factorisation scale choice, $\mh+\mz$.
Comparing the middle panels of Fig.~\ref{fig:pthz-plot}, it might seem
that the choice of a uniform reweighting provides a better description
of the hard part of $p_{t,\HZ{}}$ distribution, but the apparent
agreement between \MCFM{} and \HZNNLOLHE{} results around 400-500 GeV
is accidental. In fact, at even higher transverse momenta the NNLO
result is above our \HZNNLOLHE{} prediction. This behaviour is
entirely due to the aforementioned difference in scale choice.
\begin{figure}
  \centering
  \label{fig:pthz-plot}
  \includegraphics[page=1,width=205pt]{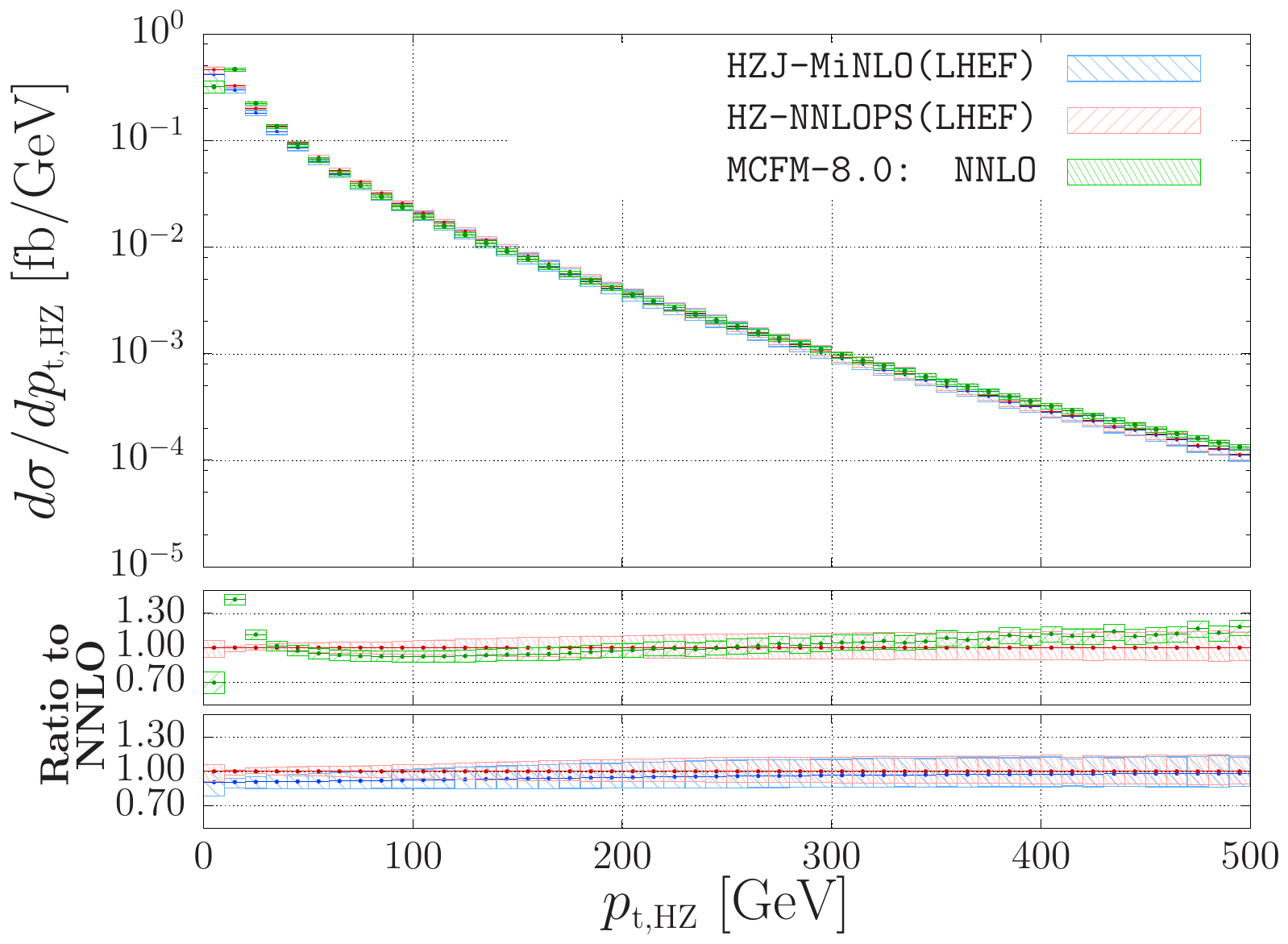}
  \hspace*{0.25cm}
  \includegraphics[page=1,width=205pt]{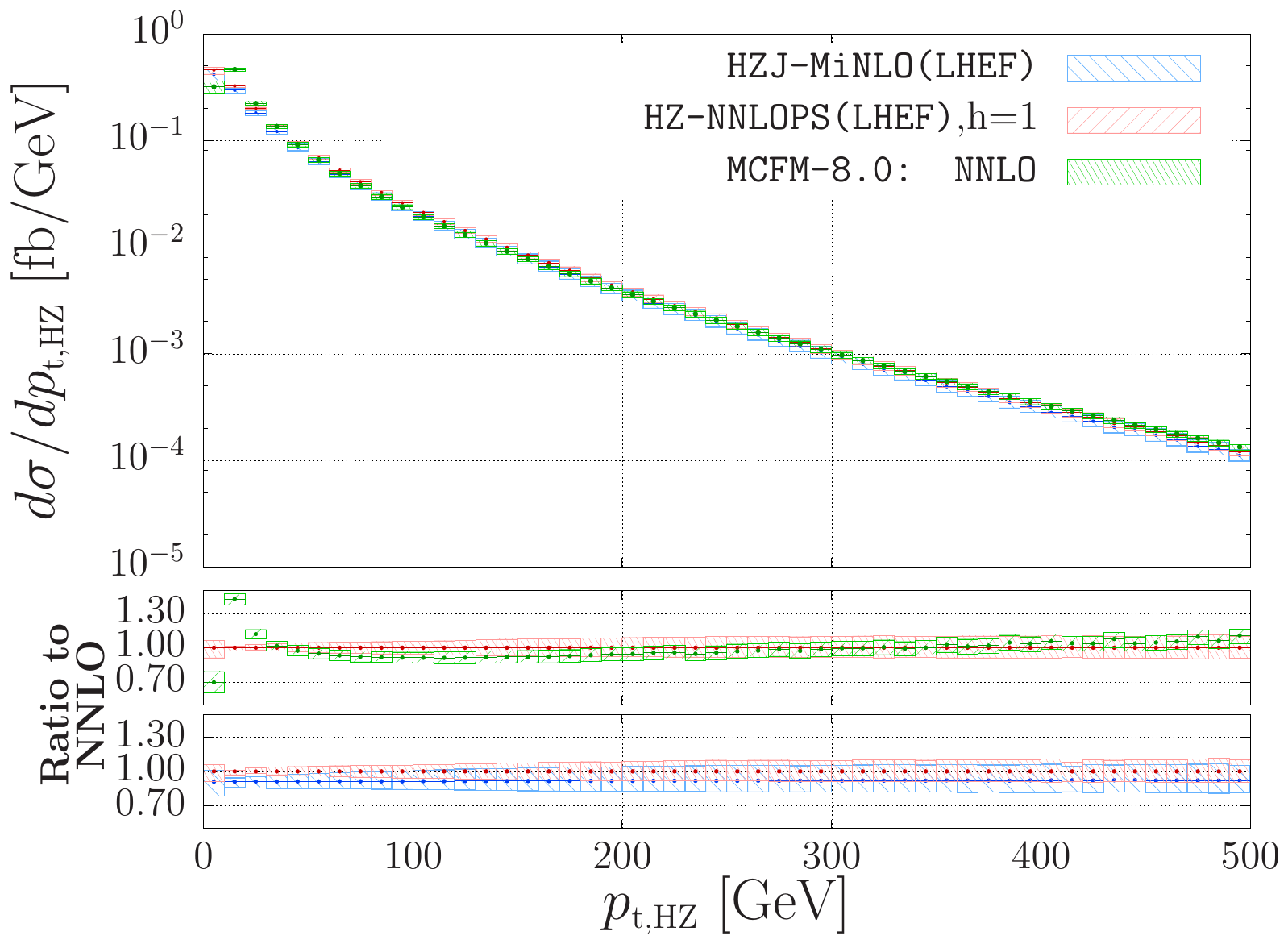}
  \caption{The differential distribution of the transverse momentum of
    the \HZ{} system results when reweighting with damping factor
    $h(p_t)$ (left panel) or without (right panel). The one-loop squared terms
    from the $\ggHZ{}$ channel have not been included.}
  \label{fig:pthz-plot}
\end{figure}

\section{Phenomenological results}
\label{sec:pheno}

In this section we turn to the discussion of the phenomenological
results obtained with our new code. We stress again that we consider
the production of a Higgs boson in association with a Z boson and
their subsequent decays $\Hbb{}$ and $\rm{Z}\to
\ell^{+}\ell^{-}$, where in the following $\ell$ denotes a single
leptonic species,~\emph{e.g.} $e$ or $\mu$.  For the Higgs boson decay we
include NLO QCD corrections.

We consider 13 TeV LHC collisions.  We consider fiducial cuts inspired
by the recent ATLAS analysis of ref.~\cite{Aaboud:2017xsd}. We require
two charged leptons with $|y_{\ell}| < 2.5$ and $p_{t,\ell} > $ 7 GeV,
moreover the harder lepton should satisfy $p_{t,\ell} > $ 27 GeV. We
impose that the invariant mass of the leptons satisfies the condition
81 GeV $ < \mll < $ 101 GeV.  Additionally we require at least
two $b$-jets with $|\eta_{j}| < 2.5$ and $p_{t,j} > $ 20 GeV.  Unless
stated otherwise, jets are defined using the flavour-$k_t$
algorithm~\cite{Banfi:2006hf} with $R=0.4$. In the flavour-$k_t$
algorithm we only consider $b$-quarks to be flavoured, and all other
light quarks to be flavourless. Using $b$-tagging, such an algorithm
can be implemented in experimental analyses.
\begin{table}
\centering
\begin{tabular}{|l|c|c|c|c|}
  \hline
\bf{Fiducial cross section}  &  \HZJMINLO{} & \MCFM{} & \HZNNLOLHE{} & \HZNNLOPS{} \\
\hline
no $\ggHZ{}$
& $6.59^{+7.2\%}_{-6.2\%}$ fb
& $7.14^{+0.5\%}_{-0.9\%}$ fb
& $7.14^{+0.3\%}_{-0.4\%}$ fb
& $6.49^{+0.8\%}_{-0.6\%}$ fb
\\
\hline
with $\ggHZ{}$
& --
& $7.92^{+2.0\%}_{-1.5\%}$ fb
& $7.90^{+2.8\%}_{-2.0\%}$ fb
& $7.16^{+3.1\%}_{-2.1\%}$ fb
\\
\hline
no $\ggHZ{}$, high-$p_{t,Z}$
& $1.13^{+5.9\%}_{-5.3\%}$ fb
& $1.21^{+0.1\%}_{-0.2\%}$ fb
& $1.21^{+0.2\%}_{-0.3\%}$ fb
& $1.13^{+1.5\%}_{-1.2\%}$ fb
\\
\hline
with $\ggHZ{}$, high-$p_{t,Z}$
& --
& $1.49^{+5.3\%}_{-4.1\%}$ fb
& $1.48^{+5.3\%}_{-4.0\%}$ fb
& $1.42^{+6.9\%}_{-5.1\%}$ fb
\\
\hline
\end{tabular}
\caption{Fiducial cross section of
  $pp\rightarrow\HZ{}\rightarrow\left(b\bar{b}\right)\left(e^{+}e^{-}\right)$
  at 13 TeV with leptonic and $b$-jet cuts. The uncertainty band
  refers to the scale variation described in the text. Numerical
  errors for each prediction are beyond the quoted digits.
   }
\label{tab:fiducial-xs}
\end{table}
The fiducial cross sections in this phase-space volume at different
levels of our simulations, are reported in
Tab.~\ref{tab:fiducial-xs}. We also present results with an additional
cut on the Z boson transverse momentum, $p_{t,\rm{Z}} > 150$ GeV,
which we refer to as high-$p_{t,\rm{Z}}$ region.

We first discuss the results without $\ggHZ{}$ contribution,
over the full range of Z boson transverse momentum reported in
the first line of the Tab.~\ref{tab:fiducial-xs}. The
\HZJMINLO{} cross section
is about 8\% smaller than the full NNLO calculation from \MCFM{}. The
difference is properly accounted for by reweighting the event sample
and the cross section of \HZNNLOLHE{} and \MCFM{} are equal to each
other within the numerical accuracy (which is at the level of the last
quoted digit). The scale uncertainty from the NLO result is reduced
from about 7\% to below 1\% at the NNLO level.
The inclusion of the $\mathcal{O}(\as^2)$ $\ggHZ{}$ channel, reported
in the second line of the table, results in further increase of the
total cross section by about 10\%. In this case, the scale uncertainty
is dominated by the new contribution, which is described only at
leading order, and increases the scale uncertainty to the level of
2-3\%. This larger scale uncertainty is somehow welcome, as a scale
uncertainty below the percent level is unlikely to reflect the true
perturbative uncertainty.  This uncertainty will be reduced by an NLO
treatment of the $\ggHZ{}$ contribution.\footnote{Note that the small
  difference in the $\ggHZ{}$ contribution in \MCFM{} and \HZNNLOLHE{}
  is due to using \LHAPDF{} or \POWHEG{} routines to perform the
  running of the coupling from $\mz$ to the central scale choice
  $\mh+\mz$.}

We now discuss the impact of the parton shower on these cross sections. As
is well known, in the presence of fiducial cuts that constrain the jet
activity, as is in the case at hand, there can be a sizeable
difference between a pure fixed-order computation and results after
applying a parton shower. This is illustrated in the last two columns
of the table.
The parton shower allows for extra QCD radiation off coloured partons
which can move the $b$-jets outside the fiducial phase-space volume,
thereby reducing the recorded cross section.  The impact of parton
shower is similar in both instances, with and without the $\ggHZ{}$ contribution, and amounts to about $10$\% reduction of the cross section in the fiducial region,
while the impact is milder, $5-7$\%, in the high-$p_{\rm t,Z}$ region.

If we now examine the results with an additional $p_{t, \rm{Z}}$ cut,
reported in the last two lines of the table, we observe a reduction of
the cross section by a factor of about 5 and in general behaviours
similar to the ones described above. One point to note is that the
impact of the $\ggHZ{}$ contribution is larger in this phase space
region, which implies also larger scale uncertainties.

\begin{figure}
  \centering
  \includegraphics[page=1,width=205pt]{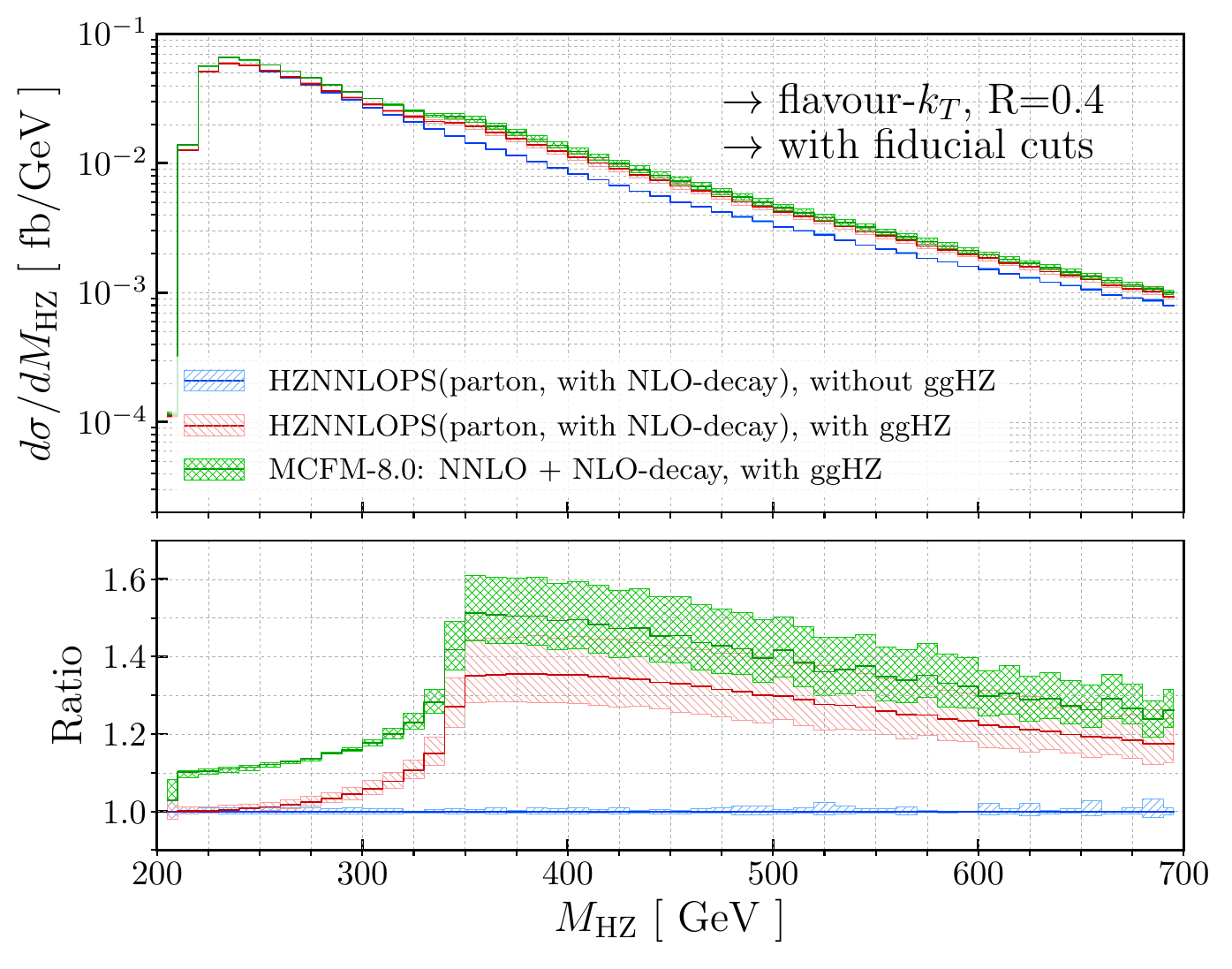}
  \hspace*{0.25cm}
  \includegraphics[page=1,width=205pt]{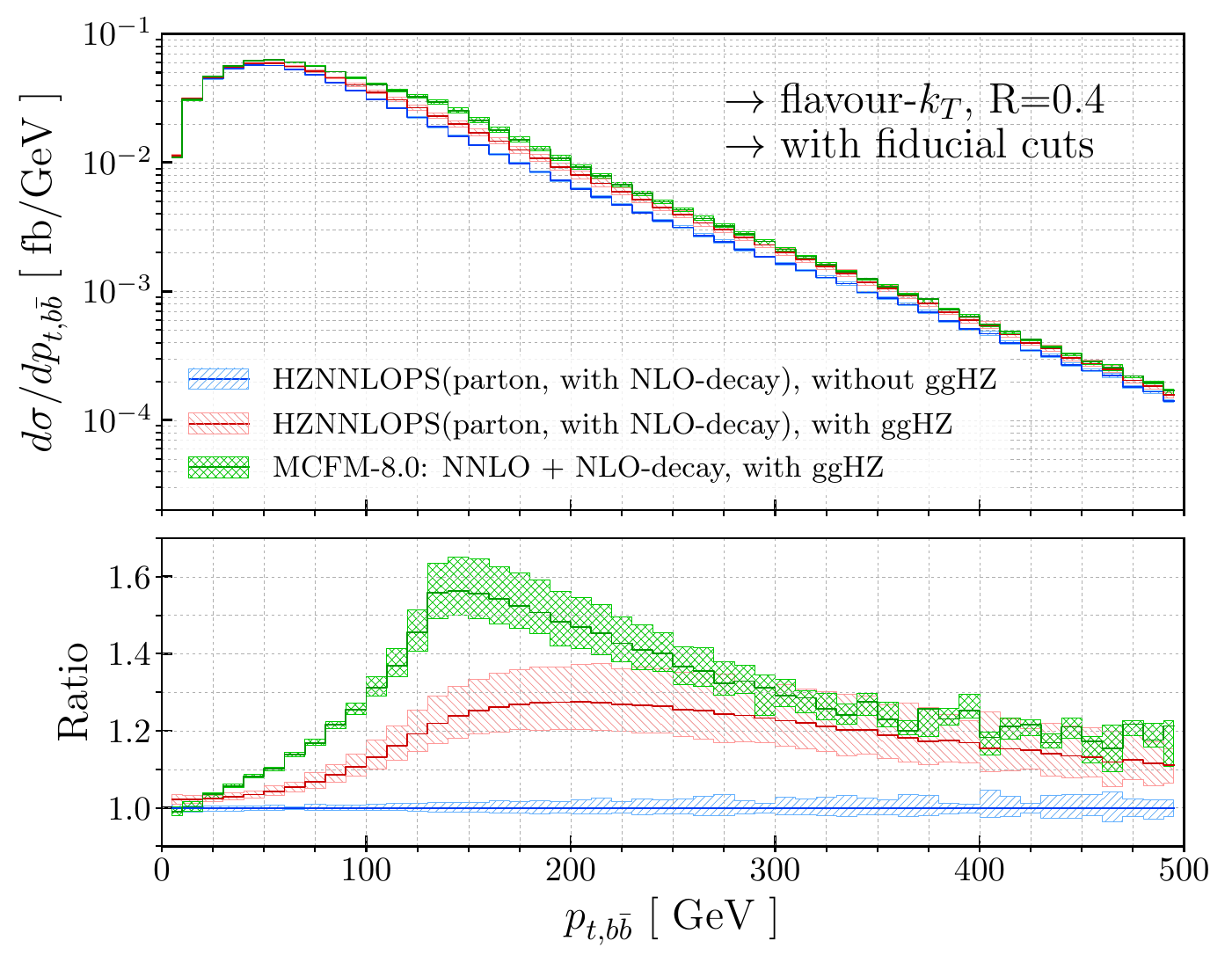}
  \caption{The differential distributions of the invariant mass of the
    \HZ{} system (left panel) and the transverse momentum of the Higgs
    boson reconstructed from $b$-jets (right panel). The lower panel
    illustrate ratio of full results (NNLO as well as NNLOPS) to the
    NNLOPS results without $\ggHZ{}$ contribution.}
  \label{fig:gghz-plots}
\end{figure}

To further illustrate the effect of the $\ggHZ{}$ channel, we present
in Fig.~\ref{fig:gghz-plots} the differential distributions of the
invariant mass of the \HZ{} system and transverse momentum of the
$b$-jet pair system ($p_{\rm t, b\bar b}$) associated to the
reconstruction of the Higgs boson momentum. To define the
\HZ{}-invariant mass we use the Monte Carlo truth, while $p_{\rm t,b
  \bar b}$ is obtained by clustering events with the flavour-$k_t$
algorithm with $R=0.4$ and by summing the transverse momenta of the
two $b$-jets.  If more than two $b$-flavoured jets are found, one
selects the pair whose invariant mass is closest to the Higgs
invariant mass.\footnote{We do not distinguish between $b$ and $\bar
  b$ jets.}
We show results from the \HZNNLOPS{} simulation before
and after the inclusion of the $\ggHZ{}$ contribution (blue and red
respectively) together with the fixed-order NNLO prediction including
the $\ggHZ{}$ contribution (green).
For the invariant mass distribution, the large impact of the $\ggHZ{}$
contribution above the top threshold is evident. Similarly, the
transverse momentum distribution is mostly affected by the $\ggHZ{}$
contribution in the region between 150 and 200 GeV. In both cases the
impact of $\ggHZ{}$ remains large up to high scales. 

The invariant mass distribution (left panel of
Fig.~\ref{fig:gghz-plots}) features an almost uniform shift between
the fixed-order predictions (green) and the ones including parton
shower evolution (red).  As discussed in the previous paragraph,
parton shower leads to a decrease in the fiducial cross section mainly
due to the $b$-jet cuts. However, the \HZ{} invariant mass is not
strictly correlated with the Higgs kinematics (and hence with the
$b$-jets from its decay).
As a consequence, we observe a moderate and constant difference
between \MCFM{} and \HZNNLOPS{}. 

On the other hand, the effect of the parton shower on the transverse
momentum of the reconstructed Higgs boson (right plot of
Fig.~\ref{fig:gghz-plots}) is quite different. In this case, one can
see that the parton shower has a sizeable effect for
$p_{t,\bb{}}\gtrsim 120$ GeV, and that it smears the distribution in
a non-uniform way. At larger values of $p_{t,\bb{}}\gtrsim 300$ GeV,
the effect of the partons shower becomes much more modest.
A more detailed discussion is presented in
App.~\ref{App:gghz}.

\begin{figure}
  \centering
  \includegraphics[page=1,width=205pt]{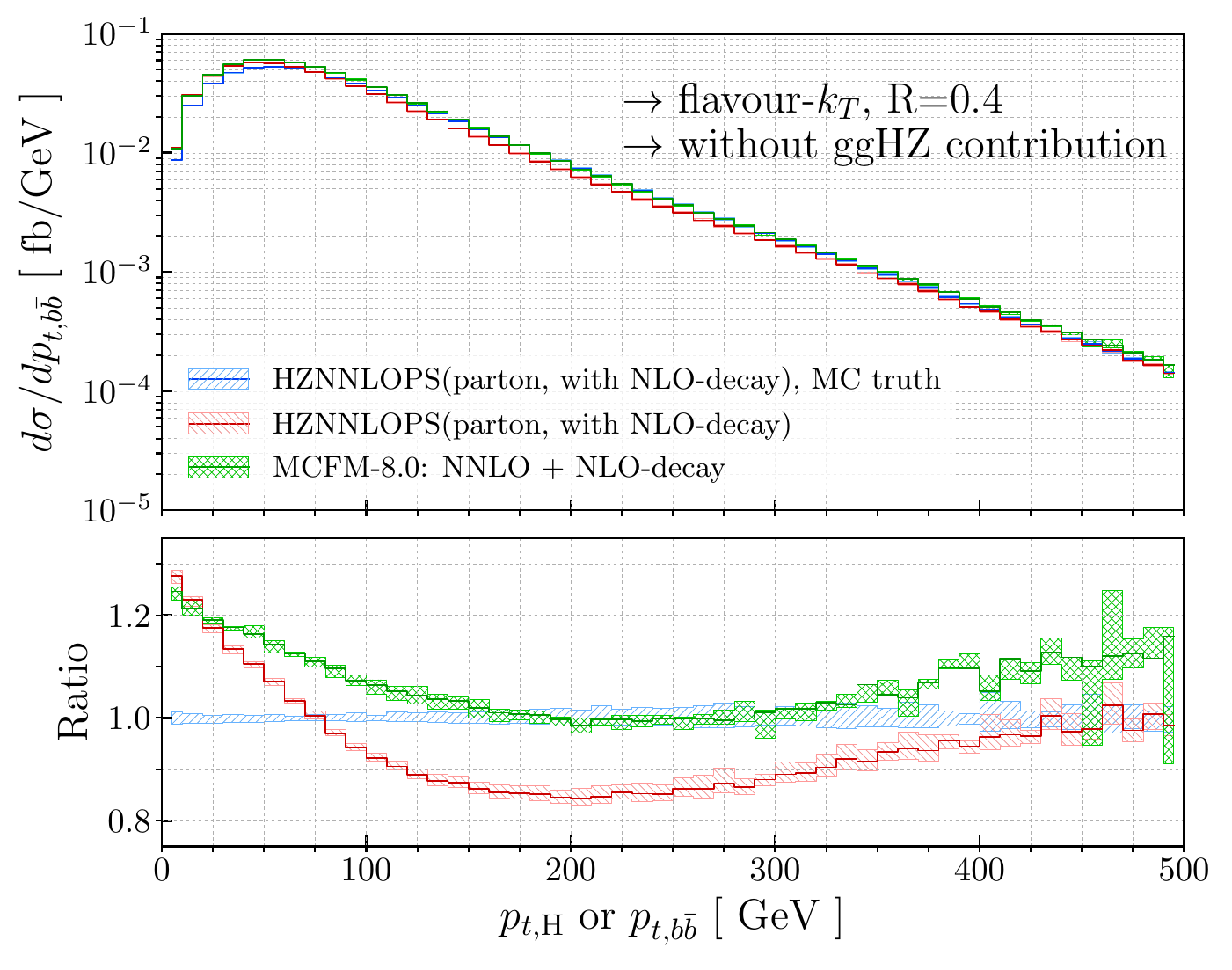}
  \hspace*{0.25cm}
  \includegraphics[page=1,width=205pt]{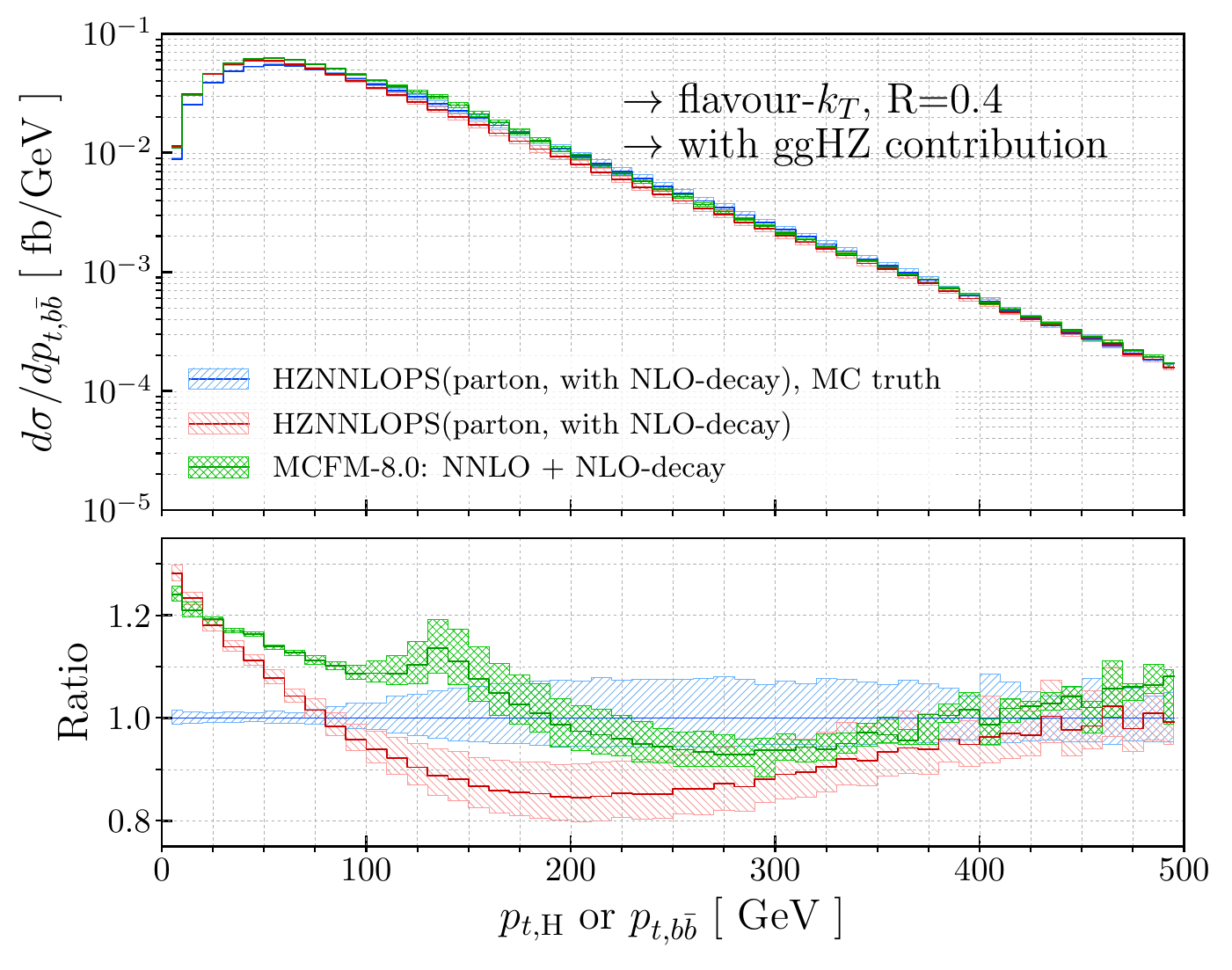}
  \includegraphics[page=1,width=205pt]{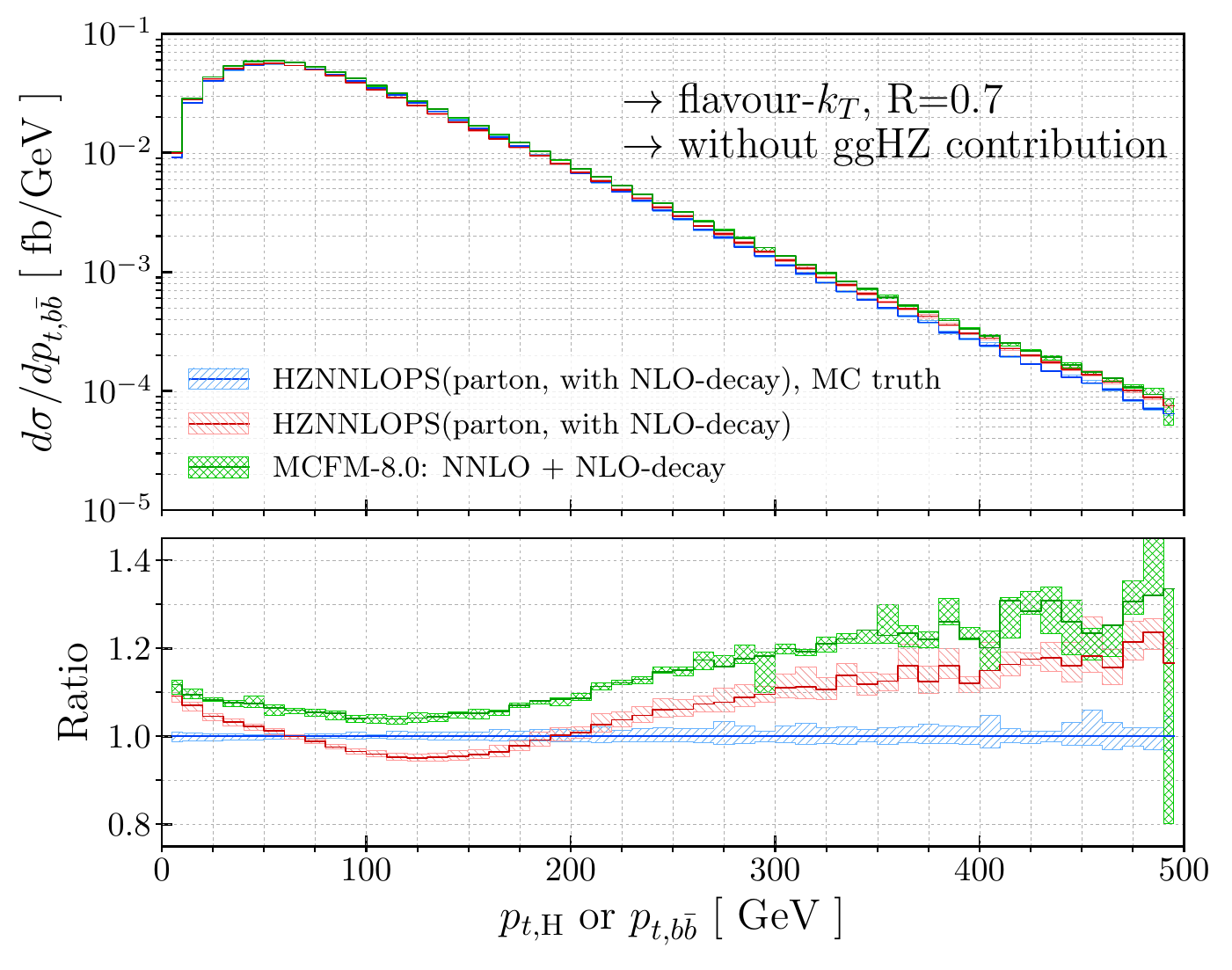}
  \hspace*{0.25cm}
  \includegraphics[page=1,width=205pt]{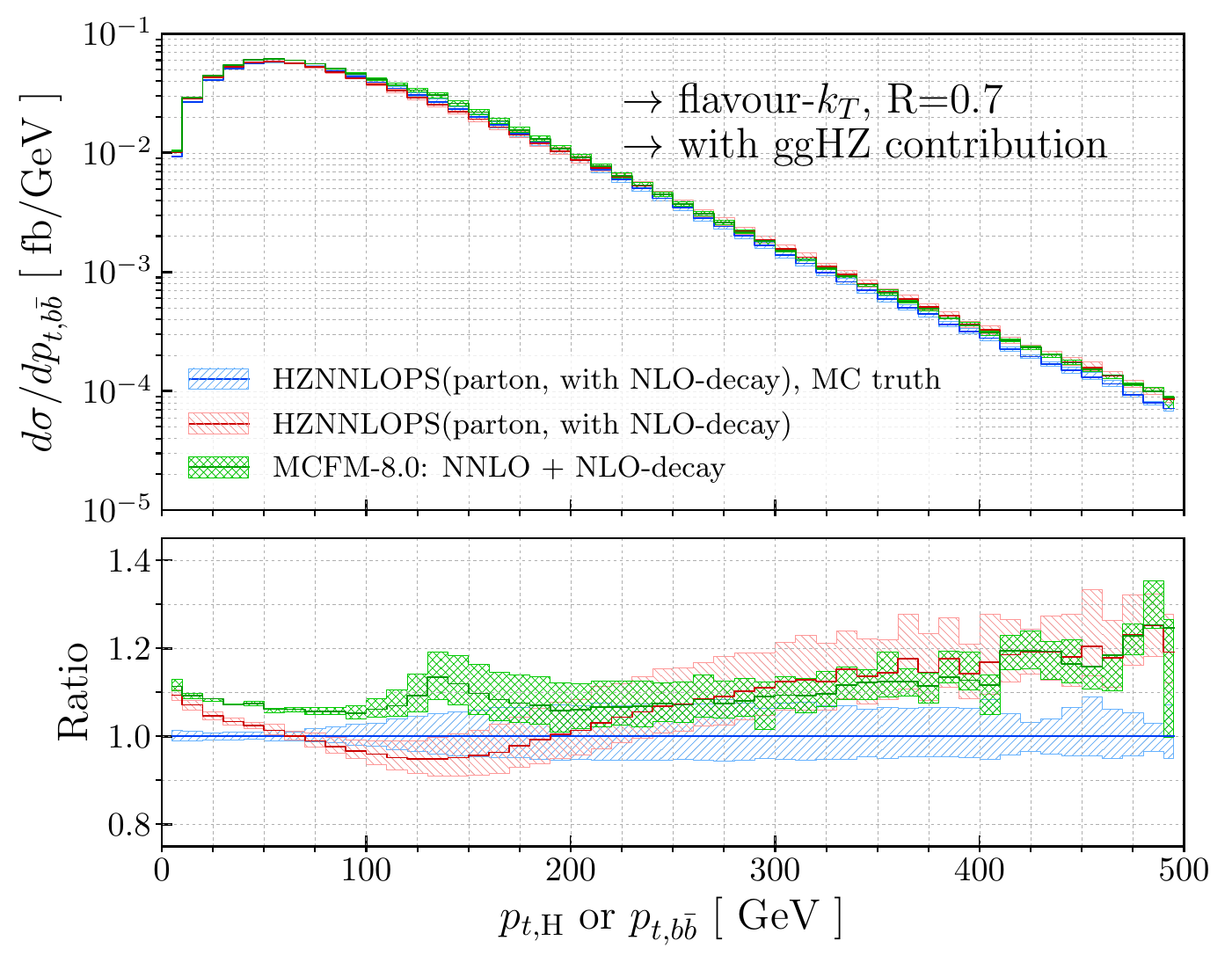}
  \caption{The differential distributions of the transverse momentum
    of the Higgs boson. \texttt{MC-truth} label refers to the actual
    Higgs boson momentum as passed from the event generator, other
    lines represent the reconstruction of the Higgs boson momentum
    using the two $b$-jets with invariant mass closest to
    $\mh{}$. The upper two plots show results for jets clustered
    with $R=0.4$, the lower plots $R=0.7$. Left plots do not include
    the $\ggHZ{}$ contribution, while the right plots do. }
    \label{fig:higgs-reconstr}
\end{figure}

As a next step, we want to study the quality of the Higgs boson
reconstruction.  In Fig.~\ref{fig:higgs-reconstr},
we present a comparison of the
transverse momentum distribution of the true Higgs boson, obtained
using the momentum passed from the event generator before it splits to
$b$-quarks and before any radiation off the $b$-quarks (labelled as
\texttt{MC-truth}), and the $\bb{}$-jet system, reconstructed with a
flavour-$k_t$ algorithm as described above. We compare two sets of plots
obtained with a jet radius of $R=0.4$ (upper plots) or $R=0.7$ (lower
plots).\footnote{We note that the fiducial cuts are applied on jets of
  $R=0.7$ in this case.} The plots show a comparison of the
    fixed-order results (green curve)\footnote{Note that here only the
      $b$-quarks from the Higgs decay are considered flavoured.}, the
    \HZNNLOPS{} after parton shower evolution (red) and the MC-truth
    prediction obtained with \HZNNLOPS{} after parton shower
    (blue). The baseline in the ratio plot is taken to be the latter.
In the left panels the $\ggHZ{}$ contribution is not included, whereas
its effect is included in right panels.

We start by examining the results without $\ggHZ{}$ contribution (left
hand panels). We note that both the fixed-order (green) and the
\HZNNLOPS{} after parton shower (red) differ from the
$\texttt{MC-truth}$ result (blue). At low transverse momenta, this
difference becomes smaller when a larger jet-radius is considered
(left bottom panel), which suggests that the dominant reason for the
difference is out-of-jet radiation from the $b\bar b$-final state. At larger
transverse momenta the difference with respect to the
$\texttt{MC-truth}$ is instead smaller at smaller jet-radius (top left
panel), which points to the fact that in this region the difference is
mainly due to radiation from the initial state.
We also notice that in the intermediate transverse momentum region the
fixed-order and \HZNNLOPS{} show sizeable differences for small jet
radius, while these differences are more moderate when using a larger
$R$. This can be easily understood from the fact that the
observable with larger $R$ is more inclusive and hence fixed-order and
parton shower results are in better agreement.

We now move to discuss the plots including the $\ggHZ{}$ effects.
First, we note that the red and green bands in the top right panel if
Fig.~\ref{fig:higgs-reconstr} are identical to the bands shown in the
right panel of Fig.~\ref{fig:gghz-plots}.
As expected when the radius becomes bigger (bottom right panel) the
fixed-order (green) and parton shower results (red) move closer to
each other, again because the observables become more inclusive. 
We also note that the uncertainty bands are now larger compared to the
results without $\ggHZ{}$ contribution. This was already observed for
the fiducial cross section and is due to the leading order
description of the $\ggHZ{}$ contribution. 

\begin{figure}
  \centering
  \includegraphics[page=1,width=205pt]{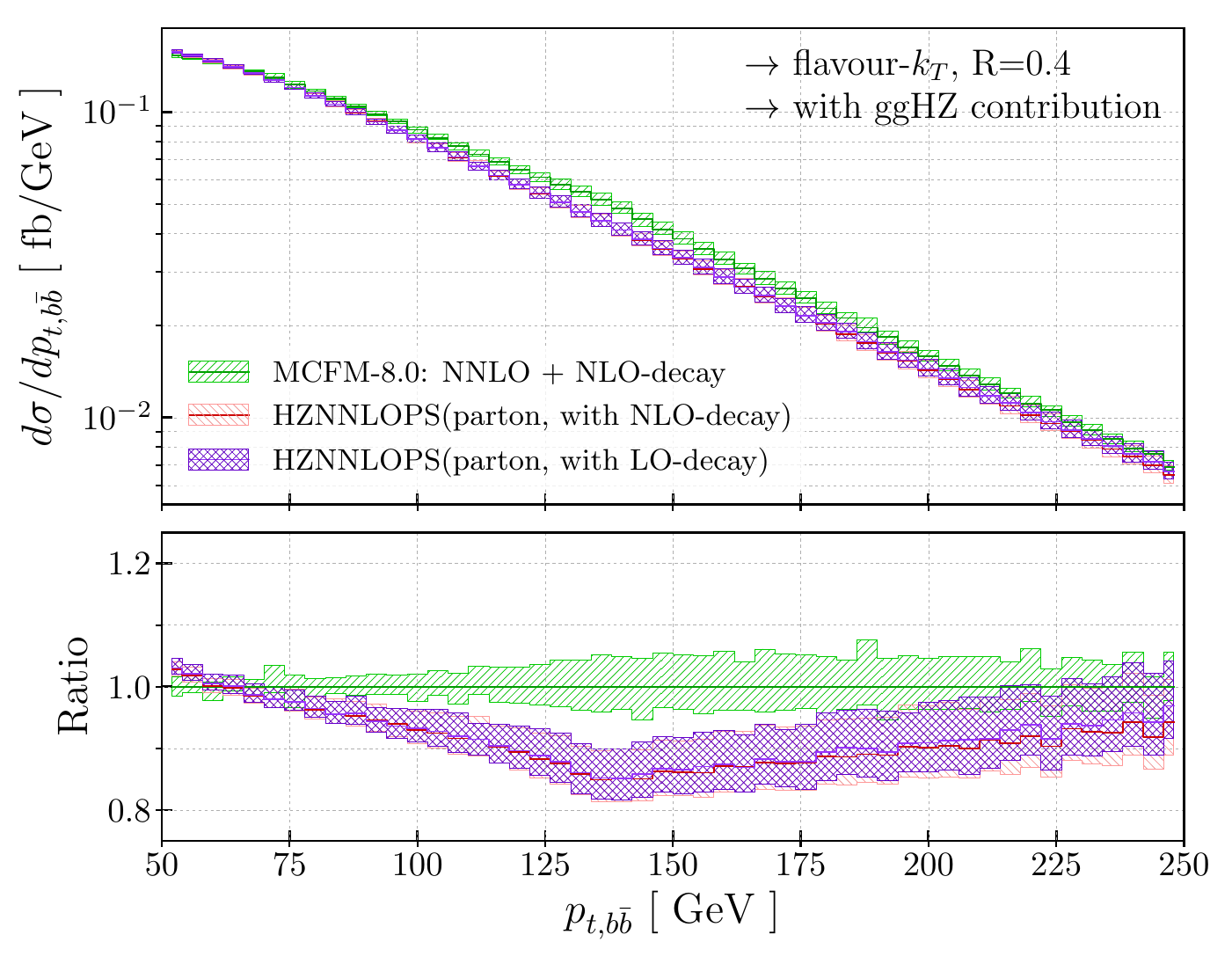}
  \hspace*{0.25cm}
  \includegraphics[page=1,width=205pt]{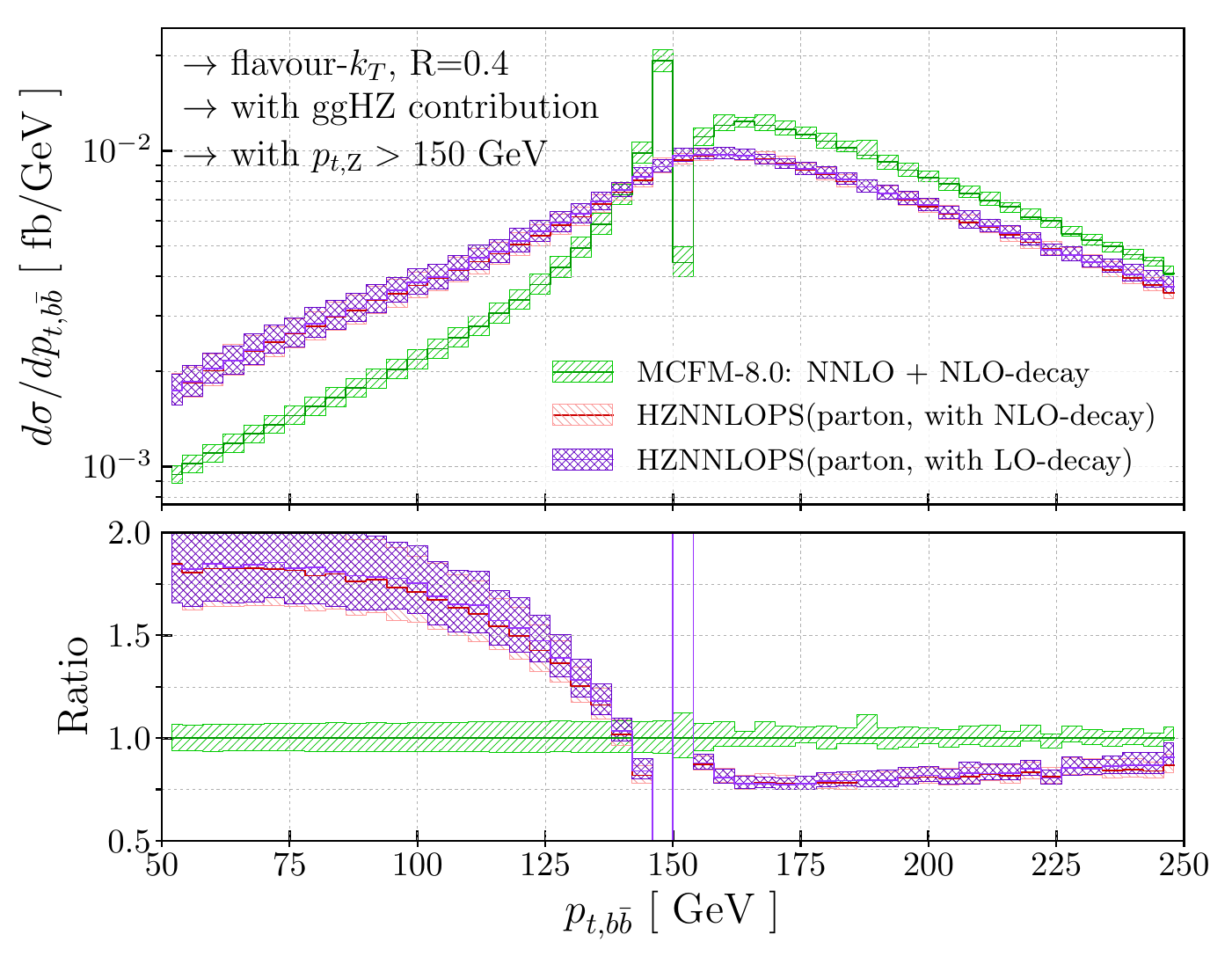}
  \caption{The differential distribution of the transverse momentum of
    the $\bb{}$-jet system without (left) and with (right) the cut
    $p_{\rm t,Z} > 150$ GeV. Results include the $\ggHZ{}$
    contribution.
  }
  \label{fig:ptbb-jets-zoom}
\end{figure}

We now show the distribution of the transverse momentum of the
$\bb{}$-jet system in the fiducial volume with and without the
additional cut $p_{\rm t,Z} > 150$ GeV. The relevant plots are shown
in Fig.~\ref{fig:ptbb-jets-zoom}. First of all we note that the
difference between treating the $\Hbb{}$ decay at NLO with respect to
LO is very small, which leads to the conclusion that a parton shower
equipped with Matrix Element corrections to the $\Hbb{}$ branching
provides a very good estimation of the higher-order corrections. We
also notice a Sudakov shoulder in the fixed-order prediction in the
right panel of Fig.~\ref{fig:ptbb-jets-zoom} at $p_{{\rm t},b\bar b} =
150$ GeV. This feature has already been observed in Figs.~(6) and (12)
of ref.~\cite{Caola:2017xuq} and is due to the fact that the presence
of the $p_{\rm t,Z} > 150$ GeV cut makes the differential spectrum
sensitive to soft gluon emission close to the cut. As expected, a parton
shower captures parts of the resummation effects and therefore the
shoulder is not present in the \NNLOPS{} predictions.

\begin{figure}
  \centering
  \includegraphics[page=1,width=205pt]{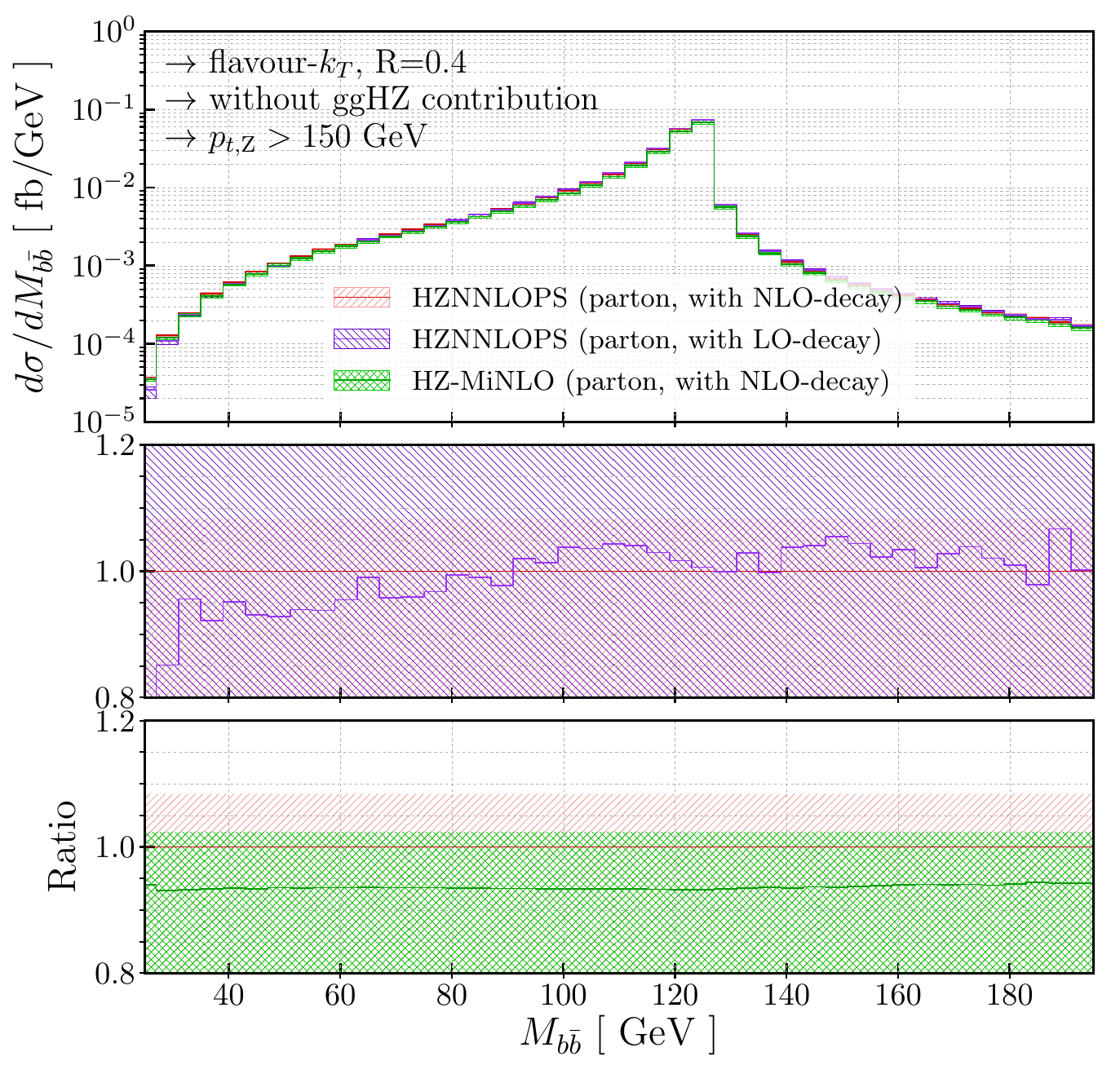}
  \includegraphics[page=1,width=205pt]{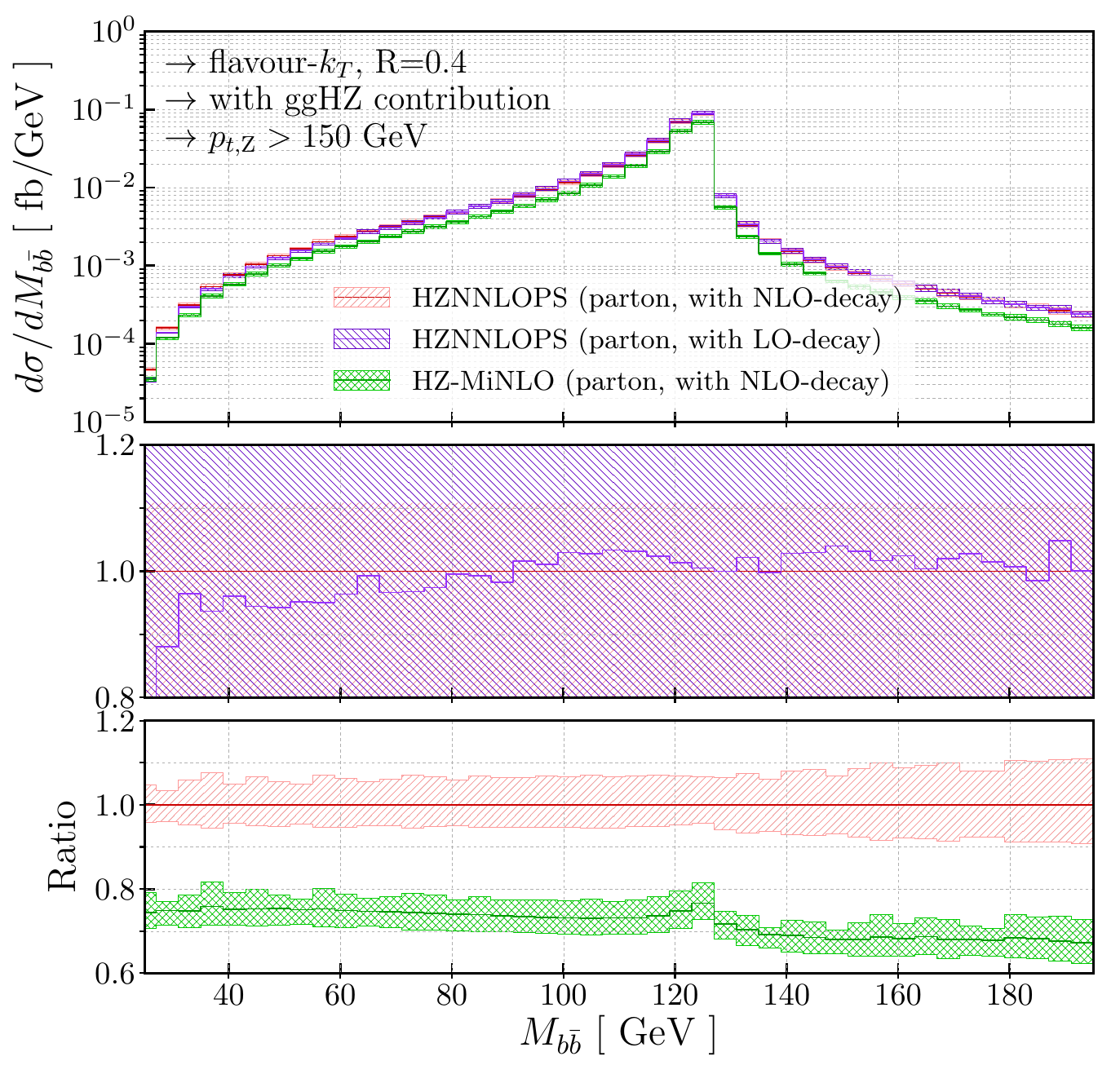}
  \caption{The differential distributions of the invariant mass of the
    $\bb{}$-system used for reconstruction of the Higgs
    boson. Comparison of \HZNNLOPS{} results with LO and NLO decay
    matrix elements, excluding $\ggHZ{}$ channel (left panels) and
    with $\ggHZ{}$ (right panels).}
  \label{fig:mbb-LOvsNLO}
\end{figure}

One of the most important variables when reconstructing a resonance is
the invariant mass of its decay products, therefore we will focus on
it in the following, in the boosted high-$p_{\rm t,Z}$ region.  
At LO in the decay the $\mbb$ distribution is an extremely narrow
Breit-Wigner function, and receives sizeable corrections away from the
peak only at higher-orders.
We start by examining how well \PYTHIA{8} can describe the decay of
the Higgs boson by comparing two calculations that include LO or NLO
decay in the matrix element. When the matrix element is computed at LO
only, \PYTHIA{8} performs the shower using with hardness scale given in
the Les Houches event file.  
This comparison is shown in Fig.~\ref{fig:mbb-LOvsNLO} without (left
plot) and with $\ggHZ$ (right plot). We compare \HZNNLOPS{} with LO
treatment of the Higgs decay (purple), \HZNNLOPS{} with NLO
corrections to the $\Hbb{}$ decay (red) and \HZJMINLO{} predictions,
with NLO decay (green).
We see that the two \HZNNLOPS{} predictions are compatible with each
other all the way down to relatively low $\mbb$ masses. We note that
the scale uncertainty is very small, of the order of 2-5\%, when no
$\ggHZ{}$ contribution is included. This uncertainty increases when
$\ggHZ{}$ events are included, since these events sit at $\mbb = \mh$
before showering. The small scale variation band is not indicative of
the true uncertainty on this distribution and is related to the fact
that \HZJMINLO{} results have been reweighted to NNLO results. In
fact, the pure \HZJMINLO{} predictions, even without $\ggHZ{}$, have a
larger uncertainty. We also note that this uncertainty is also somehow
underestimated as the band does not cover the \HZNNLOPS{}
results. This is related to the well known fact that, in a plain
\POWHEG{} simulation, the scale is varied at the level of the $\bar B$
function, which is by definition inclusive over radiation, whereas the
$\mbb$ spectrum is sensitive to radiation.

\begin{figure}
  \centering
  \includegraphics[page=1,width=205pt]{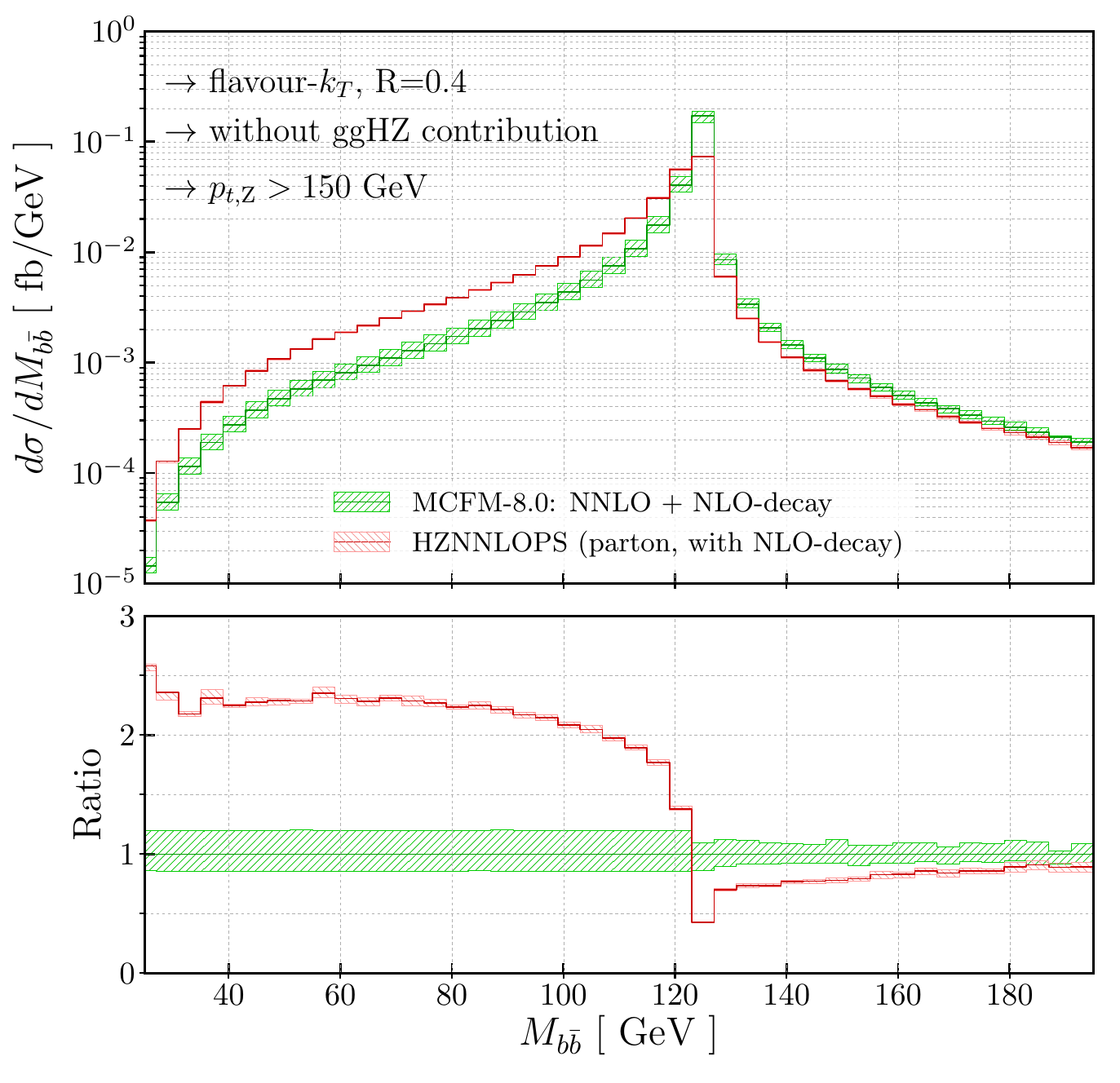}
  \hspace*{0.25cm}
  \includegraphics[page=1,width=205pt]{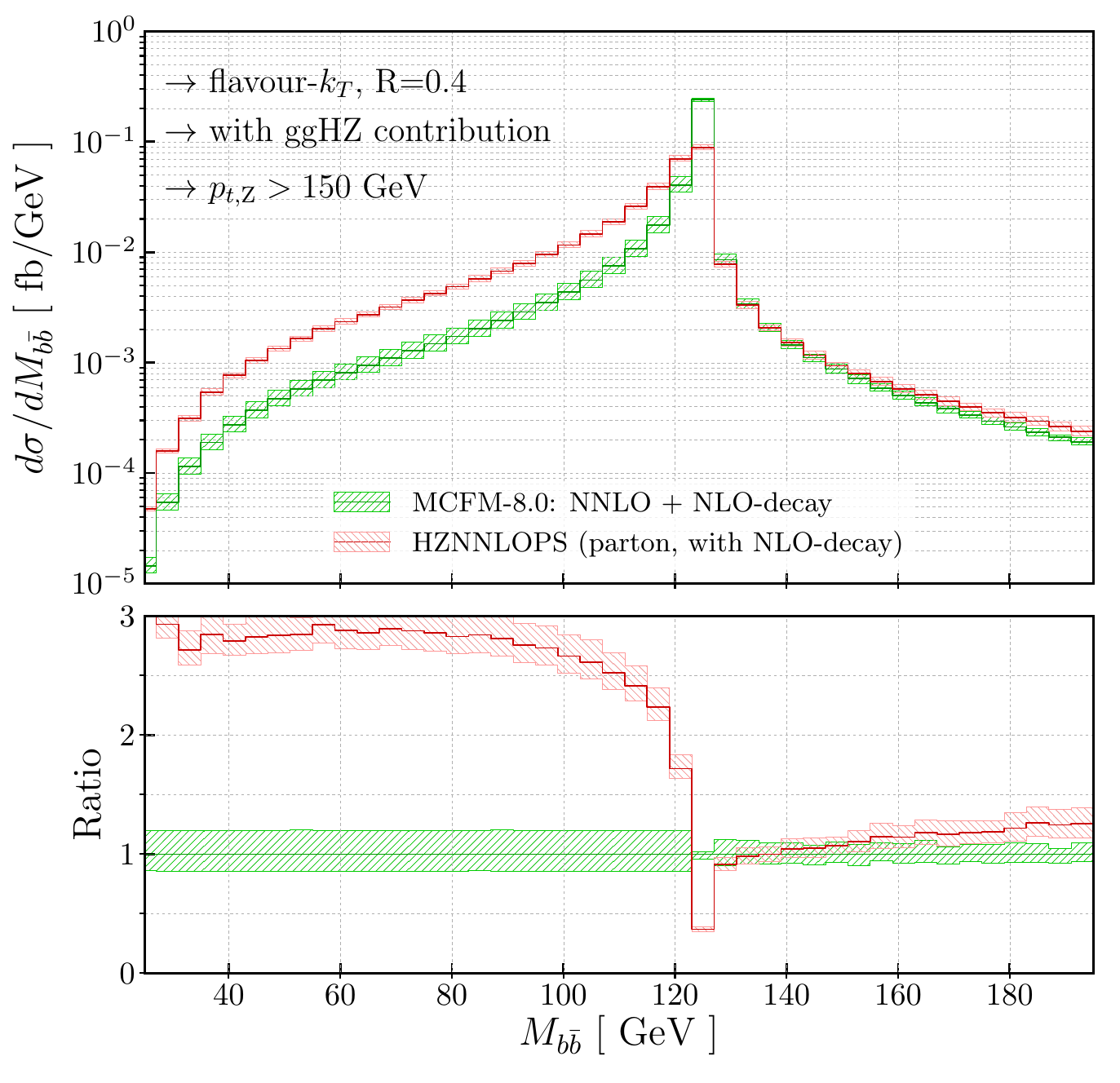}
  \includegraphics[page=1,width=205pt]{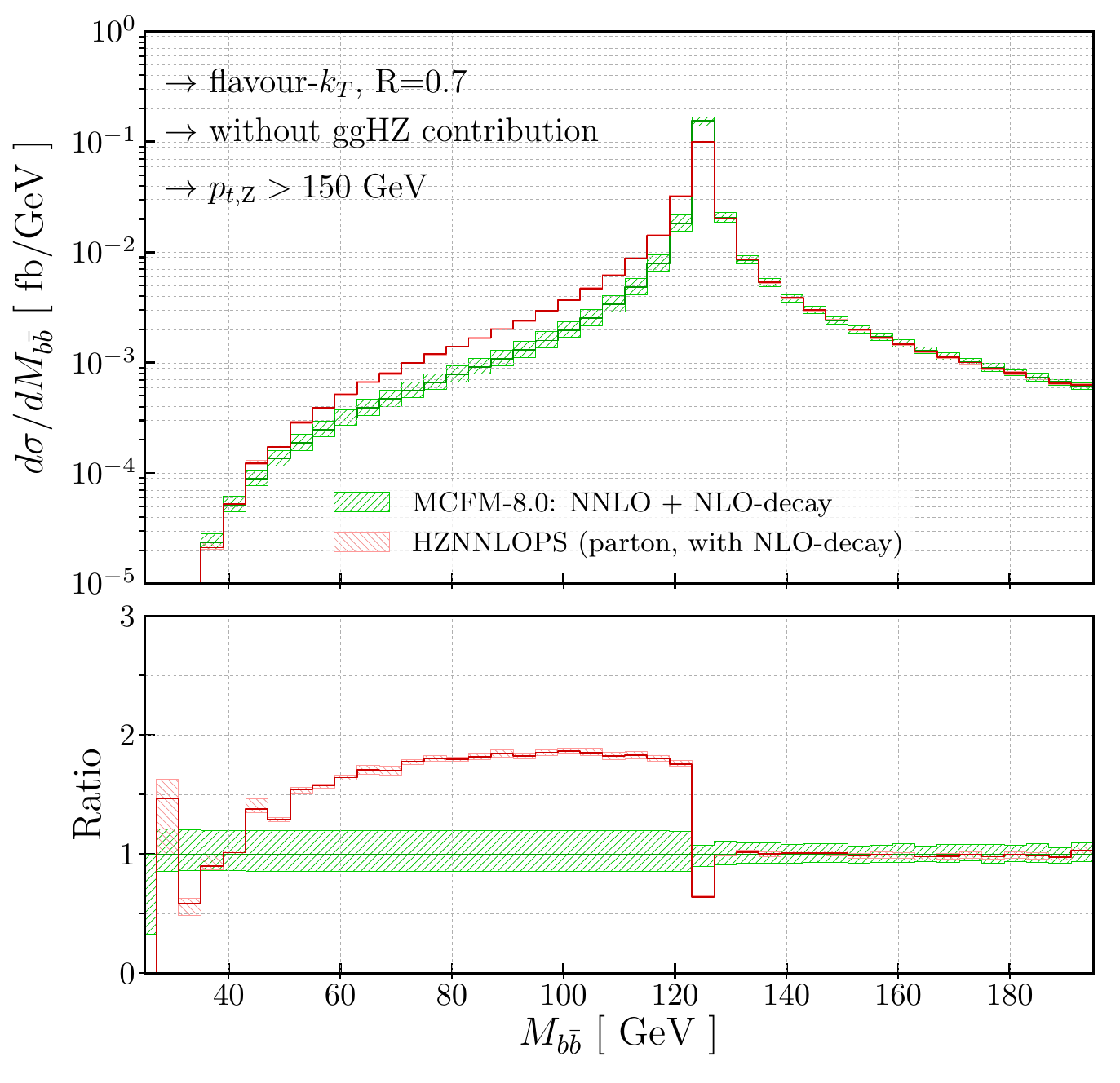}
  \hspace*{0.25cm}
  \includegraphics[page=1,width=205pt]{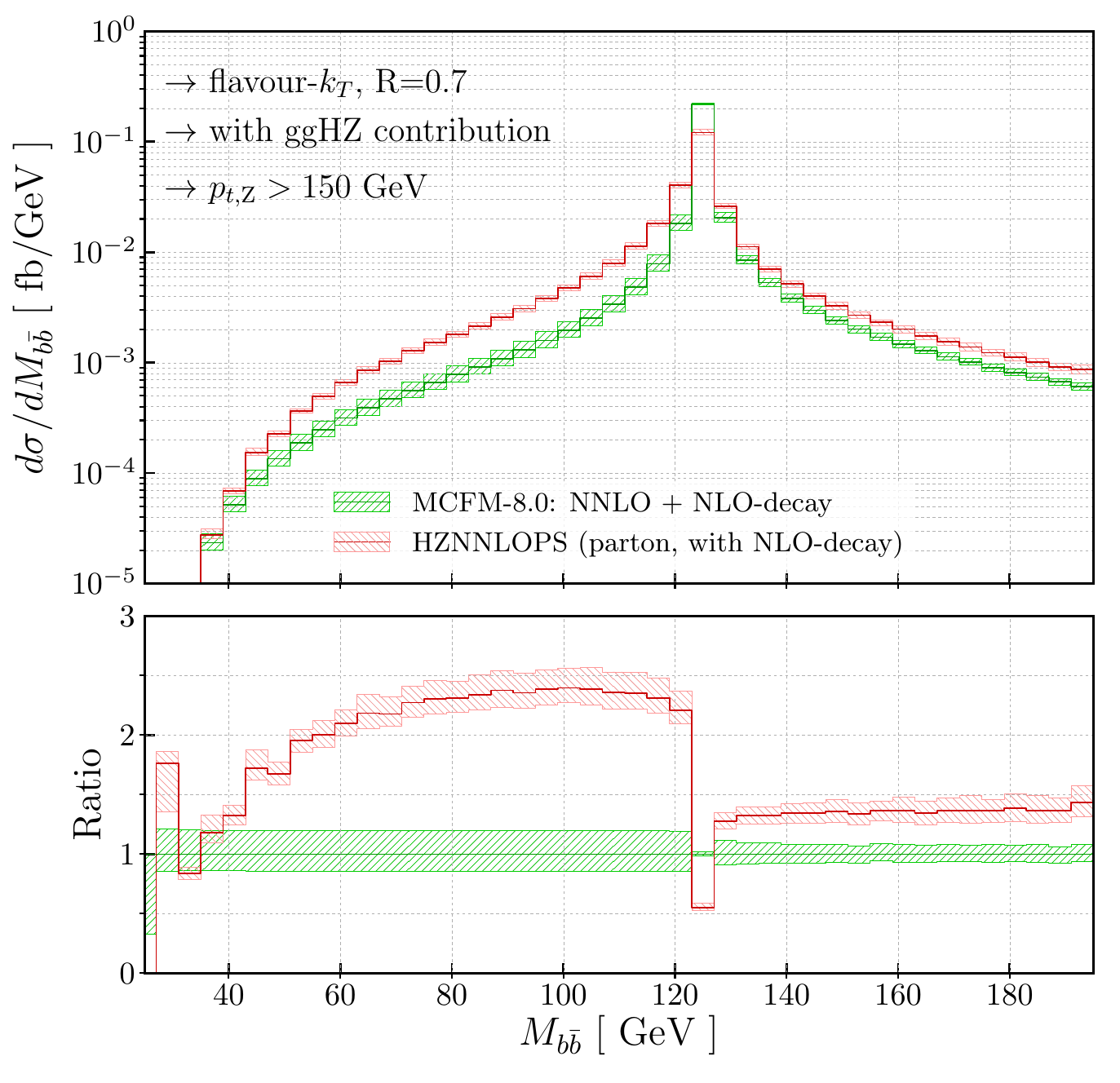}
  \caption{The differential distributions of the invariant mass of the
    $\bb{}$-system used for reconstruction of the Higgs boson. We
    present the results obtained with jet clustering with $R=0.4$ (top) and $R=0.7$ (bottom) 
    excluding $\ggHZ{}$ channel (left panels) and with $\ggHZ{}$ (right panels).}
  \label{fig:mbb-plots}
\end{figure}

In Fig.~\ref{fig:mbb-plots} we now compare fixed-order predictions
(green) and our best prediction \HZNNLOPS{} with NLO corrections to
the $\Hbb{}$ decay (red).
In the plots of Fig.~\ref{fig:mbb-plots} we
show predictions obtained with $b$-jets clustered with $R=0.4$ (top
panels) and $R=0.7$ (bottom panels).
We point out that in order to populate the region to the left of the
peak ($\mbb < \mh$) there must be a radiation off the
$b$-quarks produced in the Higgs decay. On the contrary, the region on
the right hand side of the peak ($\mbb > \mh$) is filled
only when additional radiation, off the partons from the production stage,
is clustered with the Higgs decay products. 

In Fig.~\ref{fig:mbb-plots} we notice a sizeable enhancement in the
$\mbb$ distribution to the left of the Higgs peak. This enhancement
was already observed in refs.~\cite{Ferrera:2017zex,Caola:2017xuq} and
is even more dramatic in this case.
If we compare our left plots to the Figs. (4) and (11) of
ref.~\cite{Caola:2017xuq} we observe a larger K-factor. However there
are a number of differences. First, the results of
ref.~\cite{Caola:2017xuq} are obtained with $R=0.5$. Second, our
\MCFM{} predictions are obtained using massive $b$-quarks, while the
NNLO-approx calculation shown in ref.~\cite{Caola:2017xuq} is obtained
with massless $b$-quarks. Furthermore, the two computations use
different fiducial cuts and in~\cite{Caola:2017xuq} the process
\HW{}$^-$ is considered, rather than \HZ{}. Last, our plots show
\HZNNLOPS{} results rather than \HZJMINLO{} ones, and from
Fig.~\ref{fig:mbb-LOvsNLO} this amounts to a further increase of the
ratio by 10\% (25\%) without (with) $\ggHZ{}$.
Similar considerations apply when comparing to Fig. (2) of
ref.~\cite{Ferrera:2017zex}. 

By looking at the plots on the right of Fig.~\ref{fig:mbb-plots}, one
can observe an even more pronounced enhancement of the \HZNNLOPS{}
over the \MCFM{} K-factor when the $\ggHZ{}$ contribution is
included. This is again due to the fact that the $\ggHZ{}$ term in the
fixed-order calculation is only in the $\mbb= \mh$ bin, while this
contribution is spread to other bins by the parton shower.
A second observation is that when a large jet radius is considered
(bottom row), more radiation is clustered in the $b$-jets. As a
consequence, the distribution vanishes faster away from $\mh$. This
effect is stronger when a parton shower is included and causes the
K-factor to be smaller to the left of the peak and to even become
close to one at very low mass.

\begin{figure}
  \centering
  \includegraphics[page=1,width=205pt]{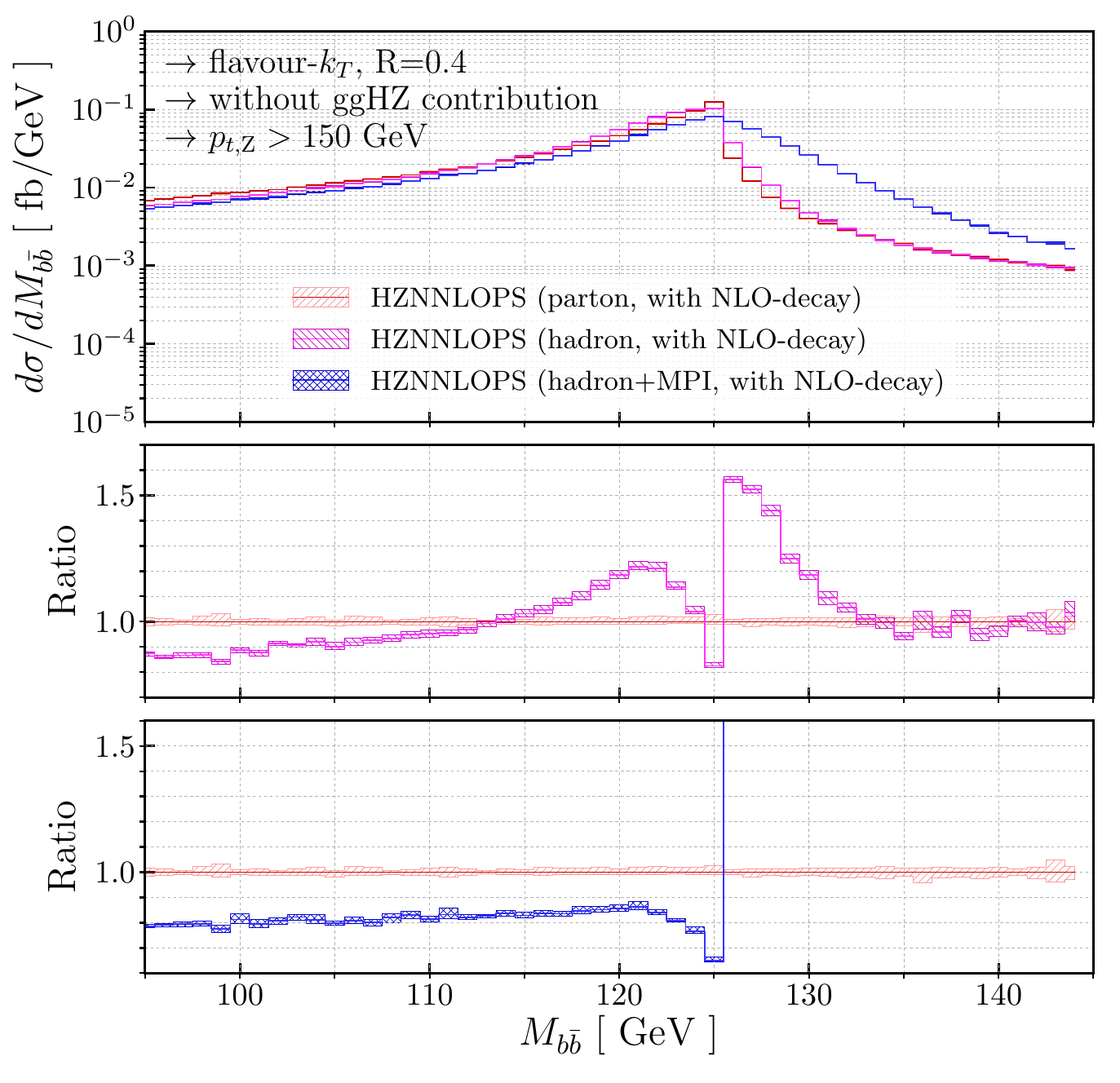}
  \includegraphics[page=1,width=205pt]{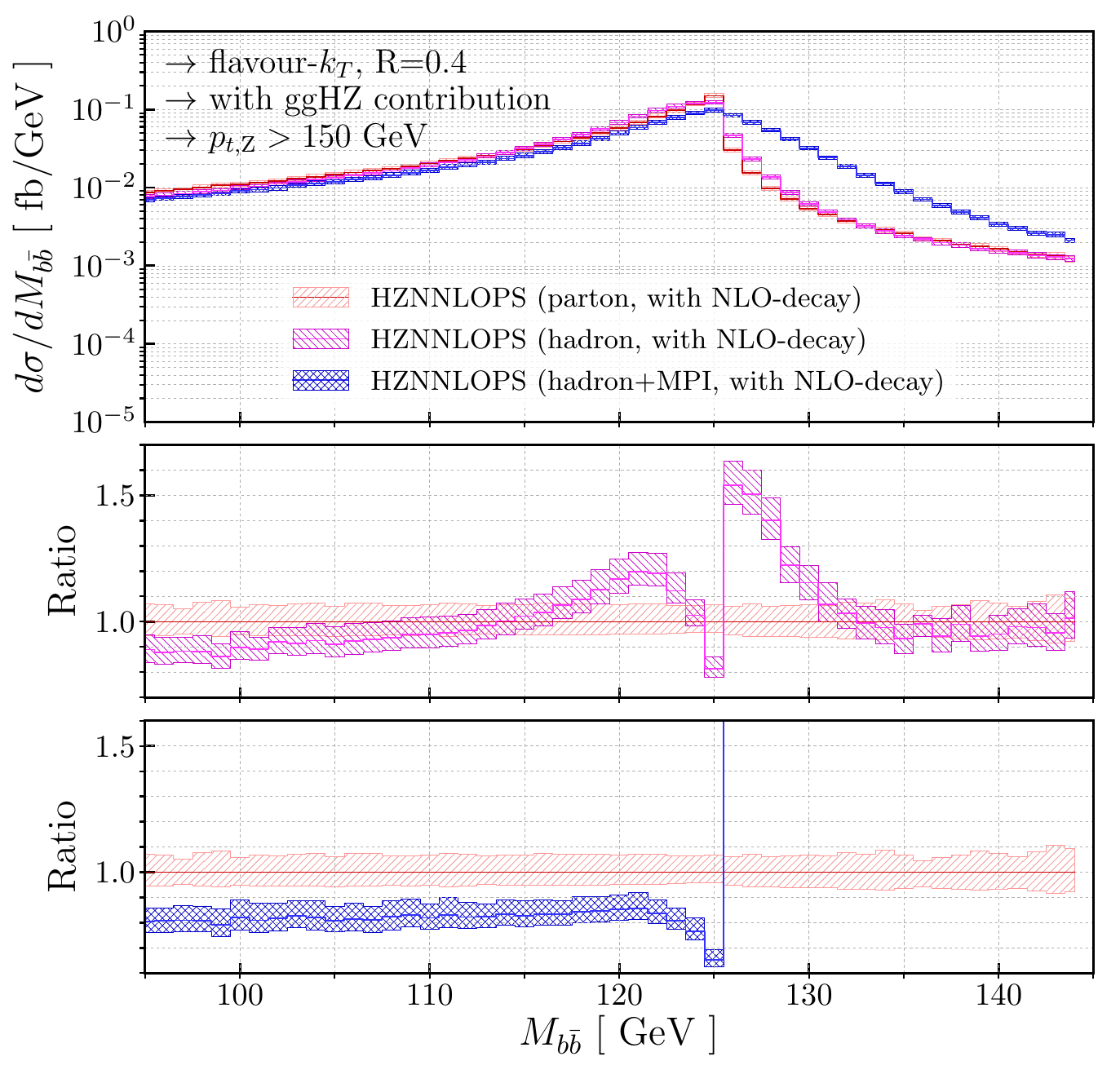}
  \caption{The differential distributions of the invariant mass of the
    $\bb{}$-system used for reconstruction of the Higgs
    boson. Comparison of \HZNNLOPS{} results with NLO decay matrix
    elements at parton-, hadron-level and with multi-parton
    interactions (MPI), excluding $\ggHZ{}$ channel (left panels) and
    with $\ggHZ{}$ (right panels).}
  \label{fig:mbb-NLO-pythia-comparison}
\end{figure}

Finally, in Fig.~\ref{fig:mbb-NLO-pythia-comparison} we present the
$\mbb$ distribution obtained with \HZNNLOPS{} at various stages after
the \PYTHIA{8} parton showering, namely at parton-level and after
hadronisation, with and without multi-parton interactions (MPI). We
notice that hadronisation smears the distribution close to the
peak. This is the reason for the dip at $\mbb=\mh$ in the first inset.
Away from the peak $\mbb=\mh$, we observe that hadronisation effects
become more important at low invariant masses, while, as expected,
they become negligible at large $\mbb$. Since color-connections tend
to reduce spatial distances between partons during hadronisation, more
radiation is clustered within the $b$-jets. This is the reason why the
small $\mbb$ regions are underpopulated with respect to predictions at
parton level. This effect is similar with or without MPI.  On the
contrary, we can see a substantial change when considering MPI in the
region $\mbb > \mh$. Since MPI provide more radiation activity, many
additional hadrons can be clustered within the $b$-jets, thereby
increasing the invariant mass of the $\bb{}$-system and causing
migration of events from the region $\mbb\approx\mh$ to larger
invariant masses.

\section{Conclusions}
\label{sec:conclu}
In this paper we have implemented a consistent matching of NNLO
accurate predictions for \HZ{} production to parton shower, including
the subsequent decay of the Z boson into pair of leptons and the NLO
decay of the Higgs boson into pair of $b$-quarks.  The \HZNNLOPS{}
generator we obtained allows for a fully-exclusive simulation of the
\HZ{} production in a hadronic collision maintaining the advantages of
the NNLO fixed-order calculation and supplying it with resummation
effects as provided by the matching to a parton shower.

In order to obtain this accuracy, we have extended the existing
\HZJMINLO{} implementation to include the NLO corrections to the
$\Hbb{}$ decay. The NNLO+PS matching procedure requires a reweighting
of the \HZJMINLO{} events that is fully-differential in the \HZ{} Born
kinematics. By using properties of the matrix elements at hand, we have
parametrised the latter using variables that allow to express the
differential cross section in terms of a finite set of functions,
thereby simplifying considerably the multi-differential reweighting
procedure. We have also included in the simulation the loop-induced
$\ggHZ{}$ channel that enters at $\mathcal{O}(\as^2)$, as it
constitutes a sizeable part of the total cross section, and can give
rise to substantial distortions of kinematic distributions.

In Sec.~\ref{sec:pheno}, we have considered a setup similar to the one
used in searches for the Higgs decay into $b$-quarks. We find that
scale uncertainties are substantially reduced when the NNLO
corrections are included. Moreover we notice that the cross section in
the fiducial region is reduced by about 5-10\% during parton shower
evolution as a consequence of requiring two $b$-jets satisfying the
fiducial cuts. This correction brings the final result outside the
NNLO uncertainty band. This highlights the limitation of using the
scale uncertainty as an estimate of the true theoretical error
associated to missing higher orders, in particular when more exclusive
fiducial cuts are applied. A NNLO+PS simulation allows to capture some
of these higher order effects, albeit with limited logarithmic
accuracy.

As already noted in the literature, we also find that the $\ggHZ{}$
channel has a significant impact, especially for the $\mhz > 2m_t$
part of the spectrum. Moreover its presence has a very strong impact
on the size of the scale uncertainty band. Since this contribution
only enters at $\mathcal{O}(\as^2)$ level, in order to reduce this
uncertainty one needs to include higher-order corrections to the
discussed channel, which are currently unknown.

We also notice differences between distributions of the transverse
momentum of the Higgs boson, computed using Monte Carlo truth, and the
transverse momentum of the $\bb{}$-system identified and used for
the reconstruction of the Higgs boson momentum. We point out that,
especially when a large jet-radius is used, the amount of radiation
clustered into $b$-jets leads to a harder $p_T$-spectrum than the one
of the true Higgs boson.

Despite the consistency of the procedure we used in our simulation, we
have obtained a large K-factor in the $\mbb{} < \mh$ region of the
distribution of the invariant mass of the $\bb{}$-system. We also
point out that the scale uncertainty of the \NNLOPS{} prediction in
this part of the spectrum is underestimated, due to known properties
of the algorithm used by \POWHEG{} to generate real radiation. These
two issues will certainly require further studies. Including NNLO
corrections to the $\Hbb{}$ decay and matching them to parton showers
would also be desirable, as well as trying to incorporate NLO
electroweak effects as obtained in
ref.~\cite{Granata:2017iod}. We leave this to future work.

\section*{Acknowledgments}
We would like to thank Gavin Salam for providing a private implementation
of the flavour-$k_t$ algorithm \cite{Banfi:2006hf}. We also are
grateful to Fabrizio Caola and Gionata Luisoni for providing some
results of ref.~\cite{Caola:2017xuq} for cross-checks and to John
Campbell, Giancarlo Ferrera, Keith Hamilton, and Paolo Nason for
useful discussions.
This work was supported in part by ERC Consolidator Grant HICCUP
(No.\ 614577). The work of ER is supported by a Maria
Sk\l{}odowska-Curie Individual Fellowship of the European Commission's
Horizon 2020 Programme under contract number 659147
PrecisionTools4LHC.
WA and WB thank CERN for their hospitality while part of this work was
done.

\appendix

\section{Treatment of the $\Hbb{}$ decay at NLO}
\label{App:higgs-nlo-decay}
The NLO corrections to the Higgs boson decay into two fermions
$f\bar{f}$ have been known for a long time
\cite{Braaten:1980yq,Drees:1990dq,Sakai:1980fa,Janot:1989jf}. We have
included them in the \HZJMINLO{} generator by extending the lists of
the flavour structures considered by the process at hand to contain
$b\bar{b}$ and $b\bar{b}g$ from the Higgs decay; creating the lists
for the corresponding resonance structures; and
by modifying the functions \texttt{setborn}, \texttt{setvirtual}, and
\texttt{setreal} to supply amplitudes for the decay.
The virtual corrections have been stripped
of infra-red and ultra-violet singularities as described in
section 2.4 of~\cite{Alioli:2010xd}. The relevant formulae read
\begin{eqnarray}
  \left| \mathcal{M}^{(0)}_{Hbb} (p_H) \right|^2
  &=&
  2\sqrt{2} N_c G_F \, p_H^2\beta^2\,m_b^{2}(\mu_r) \,,
\end{eqnarray}
\begin{eqnarray}
  \label{eq:Hbb-virt}
  2\Re\left( \mathcal{M}^{(0)}_{Hbb} (p_H)^{*} \mathcal{M}^{(1)}_{Hbb} (p_H)\right)
  &=& \left| \mathcal{M}^{(0)}_{Hbb} (p_H)\right|^2 C_F \left(\frac{\alpha_s}{2\pi}\right)\nonumber \\
  & &\hspace{-3.5cm}
  \times\Bigg\{
    -2 - 4\log{(2)}
    + \frac{1+\beta^2}{\beta} \left(\frac{2\pi^2}{3} + 2\,\rm{Li}_{2}(\xi) \right)
    \notag\\
    & &
    \hspace{-3cm}\quad
    + \frac{1+\beta^2}{2\beta}\log(\xi)\Big[ \log(\xi) + 4\log(\beta) \Big]
    \notag\\
    & &
    \hspace{-3cm}\quad
    + \frac{1-\beta^2}{\beta}\Big[ -2\log(\xi) \Big] + 2\Big[ \log(1-\beta) + \log(1+\beta) \Big]
    \notag\\
    & &
    \hspace{-3cm}\quad
    -\Big[ \frac{1+\beta^2}{\beta}\log(\xi) + 2 \Big]\log{\left(\frac{\mu_r^2}{p_H^2}\right)}
    + \left[ -3\log{\left(\frac{m_b^2}{\mu_r^2}\right)} + 4 \right]
    \Bigg\}\,,
\end{eqnarray}
\begin{eqnarray}
  \left| \mathcal{M}^{(0)}_{Hbbg} (p_H,q_1,q_2)\right|^2
  &=&
  \left| \mathcal{M}^{(0)}_{Hbb} (p_H)\right|^2 C_F \left(\frac{\alpha_s}{2\pi}\right) \cdot\frac{4 \pi^2}{\mhsq{}\,\beta^2}
  \Bigg[ 8 + 4\frac{\qk{1}}{\qk{2}} + 4\frac{\qk{2}}{\qk{1}} \notag\\
    & & +\beta^2 \left( \frac{\mhsq}{\qk{1}}\,\frac{\mhsq}{\qk{2}}
    - \frac{1}{2}\,\left(\frac{\mhsq}{\qk{1}}\right)^2
    - \frac{1}{2}\,\left(\frac{\mhsq}{\qk{2}}\right)^2
    - 4\,\frac{\mhsq}{\qk{1}}
    - 4\,\frac{\mhsq}{\qk{2}}
    \right)\notag\\
    & &+\beta^4\left(
      \frac{1}{2}\,\left(\frac{\mhsq}{\qk{1}}\right)^2
      + \frac{1}{2}\,\left(\frac{\mhsq}{\qk{2}}\right)^2
      + \frac{\mhsq}{\qk{1}}\,\frac{\mhsq}{\qk{2}}
    \right)
    \Bigg]\,,
\end{eqnarray}
where $p_H$ is the momentum of Higgs boson, $q_1$ and $q_2$ are the
momenta of the $b$ and $\bar{b}$ quarks respectively, $k$ is the
momentum of the gluon in the real radiation matrix element,
\begin{equation}
  x_b^{\,2}= \frac{m_b^2}{p_H^2}\,,
  \quad
  \beta = \sqrt{1-4 x_b^2}\,,
  \quad
  {\rm{ and }}
  \quad
  \xi = \frac{1-\beta}{1+\beta}
  \,.
\end{equation}
With $m_b(\mu_r)$ we denote the $b$-quark mass in the $\MSbar$ scheme,
evaluated at the decay renormalisation scale $\mu_r$. For the case at
hand, we pick the Higgs boson mass as the central value for $\mu_r$
and its variation is correlated with the production renormalisation
scale variation,~\emph{i.e.} we use the same scaling factor for
$\mu_r$ and $\mur$ (where the latter is the renormalisation scale used
for the production matrix elements, as introduced in
Sec.~\ref{sec:hbb-nlo-dec}). The last term in Eq.~\eqref{eq:Hbb-virt}
denotes a change from on-shell scheme to $\MSbar$ scheme, namely using
Eq.~(63) of ref.~\cite{Campbell:2002zm} and retaining terms up to
$\mathcal{O}(\as)$.

\section{Spectral decomposition of polar angle distributions}
\label{App:GramSchmidt}
It is natural to use a spectral decomposition in terms of Fourier
modes when dealing with an angular variable. The polar angle $\alpha$
that we use to parametrise the kinematics is defined on the interval
[$0;\pi$]. The most generic parametrisation of the angular dependence
can be written in terms of polynomials of $\cos\alpha$ and
$\sin\alpha$ (note that quadratic and higher terms in $\sin\alpha$ can
always be reduced to at most linear piece):
\begin{equation}
  \label{eq:spectral-decomp-cos-sin}
  F(\alpha) = \sum_{a=0}^\infty \bigg(C_{1,a} (\cos\alpha)^a + C_{2,a} (\sin\alpha)(\cos\alpha)^a\bigg)\,,
\end{equation}
which equivalently reads
\begin{eqnarray}
  \label{eq:spectral-decomp-x}
  F(x)
  &=& \sum_{a=0}^\infty \bigg(C_{1,a} x^a + C_{2,a} \left(\sqrt{1-x^2}\right)x^a \bigg)\notag\\
  &=& \sum_{i=0}^{\infty} \bar{C}_{i} f_{i}(x)
\end{eqnarray}
with
\begin{eqnarray}
  f_{i}(x)
  &=& \left(\sqrt{1-x^2}\right)^{\rm{mod}(i,2)}x^{\floor{i/2}}\,, \notag\\
  x &=& \cos\alpha, \qquad x\in[-1;+1].
\end{eqnarray}
The above $f_i$ functions are not orthonormal.
Equipped with a scalar product between two functions
\begin{equation}
  \langle F | G \rangle \equiv \int_{-1}^{+1} F(x)\,G(x)\,dx\,,
\end{equation}
we can transform the basis of functions $\{f_i\}$ into an orthonormal set
$\{g_j\}$ by means of a Gram-Schmidt recurrence relation
\begin{eqnarray}
  k=0: & &\begin{cases}
    \tilde{g}_0 =& f_0, \\
    g_0 =& \tilde{g}_0 / \langle \tilde g_0 | \tilde g_0 \rangle
  \end{cases}\notag\\
  k>0: & &\begin{cases}
    \tilde{g}_{k} =& f_{k} - \sum_{j=0}^{k-1} {\langle g_j | f_k \rangle} \cdot g_j,\\
    g_{k} =& \tilde{g}_{k} / \langle \tilde g_{k} | \tilde g_{k} \rangle
    \end{cases}
\label{eq:gdef}
\end{eqnarray}
and express a generic function $F$ in terms of the $\{g_j\}$ basis
\begin{equation}
  F(x) = \sum_{j=0}^{N} c_{j} \, g_j(x)\, ,\qquad {\rm with}\qquad c_n = \langle g_n | F \rangle.
\end{equation}
For the case at hand, due to the arguments given in
App.~\ref{App:HadronicTensor}, only terms up to $N=10$ are needed.

\section{Hadronic tensor approach to matrix element}
\label{App:HadronicTensor}
Hadronic collisions of protons are inherently linked with
non-perturbative aspects of strong interactions through proton parton
distribution functions (PDFs). Nevertheless we can use Lorentz
symmetries and gauge invariance to predict the tensor structures that
appear in matrix elements for associated Higgs production in $pp$
collision. We distinguish two stages of the process: the production of
the off-shell gauge boson in hadronic collision, that may be
parametrised by the hadronic tensor, $H_{\mu\nu}$, and decay of the
gauge boson into the Higgs boson and a pair of leptons, described by
the decay tensor, $D_{\mu\nu}$. Since we are considering only QCD
corrections and do not consider interference effects between
production and decay products of the Higgs boson, the full squared
matrix element can be written as
\begin{eqnarray}
  \label{eq:hadroni-leptonic-tensor-formula}
  \left| \mathcal{M}(p_1,p_2,q,\ell_1,\ell_2)\right|^2
  &=&
  \frac{
  H_{\mu\nu}(p_1,p_2,q)
  \cdot
  D^{\mu\nu}(q,\ell_1,\ell_2)
  }
  {\left(q^2 - \mzsq{}\right)^2 + \mzsq{}\gzsq{}},
\end{eqnarray}
where $p_1, p_2$ are the momenta of the incoming protons, $q$ is the
momentum of the off-shell gauge boson before radiating off the Higgs
boson, while $\ell_1$ and $\ell_2$ are the momenta of the two leptons
that are produced. The momentum of the Higgs boson $p_H$ may be
obtained from conservation of momentum $p_H = q - \ell_1 - \ell_2$.
\begin{figure}
  \centering
  \label{fig:hadroni-tenor}
  \includegraphics[width=200pt]{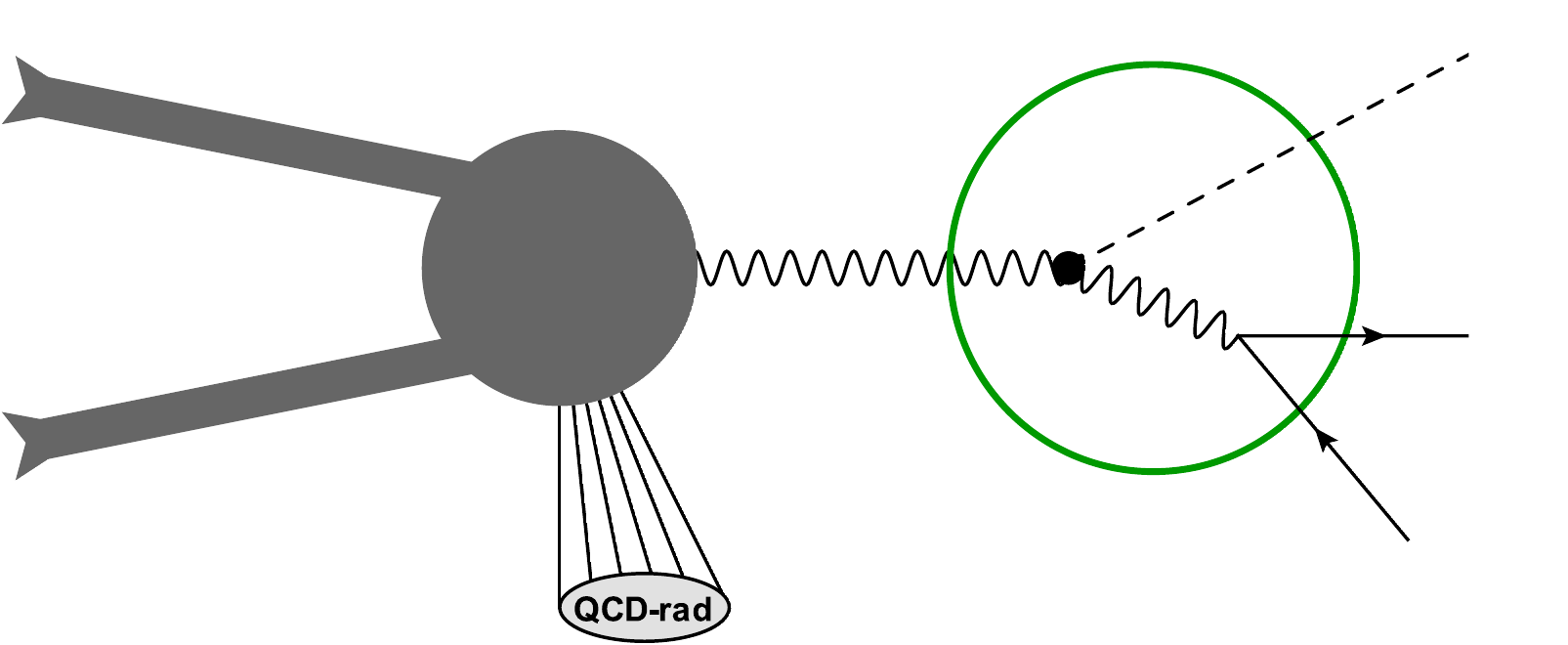}
  \caption{Amplitude for production of a weak gauge boson in
    proton-proton collision with subsequent branchings into Higgs boson and pair of leptons.}
  \label{fig:hadronic-tensor-approach}
\end{figure}
We can parametrise the hadronic tensor as
\begin{eqnarray}
  H_{\sigma\sigma'}
  =
  \left(\varepsilon(q)\right)^{\mu}_{\sigma}\,
  H_{\mu\nu}(p_1,p_2,q)\,
  \left(\varepsilon^{*}(q)\right)^{\nu}_{\sigma'},
\end{eqnarray}
where the $\varepsilon$-four-vectors denote polarisation tensors of the gauge boson
in the amplitude and its conjugate part, corresponding to polarisations
$\sigma$ and $\sigma'$ respectively.
The most general covariant form for the hadronic tensor~\cite{Korner:1990im,Lam:1978pu} reads
\begin{eqnarray}
    H_{\mu\nu}(p_1,p_2,q)
  &=& \HH{1} \left( g_{\mu\nu} - \frac{q_{\mu}q_{\nu}}{q^2}\right)
   +  \HH{2} \,\tilde{p}_{1\mu} \tilde{p}_{1\nu}
   +  \HH{3} \,\tilde{p}_{2\mu} \tilde{p}_{2\nu} \notag\\
  &+& \HH{4} \big(\tilde{p}_{1\mu} \tilde{p}_{2\nu} + \tilde{p}_{2\mu} \tilde{p}_{1\nu} \big)
   +  \HH{5} \big(\tilde{p}_{1\mu} \tilde{p}_{2\nu} - \tilde{p}_{2\mu} \tilde{p}_{1\nu} \big) \notag\\
  &+& \HH{6} \epsilon(\mu \nu p_1 q)
   +  \HH{7} \epsilon(\mu \nu p_2 q) \notag\\
  &+& \HH{8} \big( \tilde{p}_{1\mu}\,\epsilon(\nu p_1 p_2 q) + \{\mu\leftrightarrow\nu\} \big)
   +  \HH{9} \big( \tilde{p}_{2\mu}\,\epsilon(\nu p_1 p_2 q) + \{\mu\leftrightarrow\nu\} \big), 
\end{eqnarray}
where $\tilde p_{i\mu} = p_{i\mu} - (p_i q)/q^2 q_\mu$.

The decay tensor, $D_{\mu\nu}$, responsible for the part of the process
that is represented inside a green circle in
Fig.~\ref{fig:hadronic-tensor-approach}, can be parametrised as
$
  D_{\sigma\sigma'}
  =
\left(\varepsilon(q)\right)_{\mu,\sigma}\,
  D^{\mu\nu}(q,\ell_1,\ell_2)\,
  \left(\varepsilon^{*}(q)\right)_{\nu,\sigma'}  $,
  where $D^{\mu\nu}(q,\ell_1,\ell_2)$ has the structure 
(for simplicity we omit overall coupling constant factors) 
\begin{eqnarray}
  \label{eq:d_munu_structure}
  D^{\mu\nu}
  &=&
  \frac{
  g^{\mu\alpha}
  \Big( -g_{\alpha\tilde\alpha} + \frac{k_{\alpha} k_{\tilde\alpha}}{\mzsq} \Big)
  L^{\tilde\alpha\tilde\beta}
  \Big( -g_{\beta\tilde\beta} + \frac{k_{\beta} k_{\tilde\beta}}{\mzsq} \Big)
  g^{\beta\nu}}
  {\left(k^2 - \mzsq{}\right)^2 + \mzsq{}\gzsq{}}
  \,,
\end{eqnarray}
where $k=\ell_1+\ell_2$ is the momentum of the gauge-boson after radiating off Higgs boson
and 
\begin{eqnarray}
  \label{eq:tr_fermion_line}
    L^{\tilde\alpha\tilde\beta}
  =
  \textrm{Tr}\left[
    \gamma^{\tilde\alpha}\left(V_{\ell}+A_{\ell}\gamma_5\right)
    \slashed{\ell_1}
    \gamma^{\tilde\beta}\left(V_{\ell}+A_{\ell}\gamma_5\right)
    \slashed{\ell_2}
    \right], 
\end{eqnarray}
where $V_l$ and $A_l$ are vector and axial components of the coupling
of the vector boson to lepton.  We can express the lepton momenta as
$\ell_1 = \ell$ and $\ell_2 = k - \ell$. Plugging this into
Eq.~\eqref{eq:tr_fermion_line} and then further into
Eq.~\eqref{eq:d_munu_structure}, we find out that the momentum $k$ of
the Z boson after radiating off the Higgs boson appears in the final
amplitude in Eq.~\eqref{eq:hadroni-leptonic-tensor-formula} with at
most power 5. This argument is analogous to the one used in
ref.~\cite{Collins:1977iv} to obtain the general form of the angular
dependence of Drell-Yan decays.

\section{Impact of the $\ggHZ{}$ contributions}
\label{App:gghz}
In this section we discuss the numerical impact of the
loop-induced $\ggHZ{}$ contribution, that we include as explained in
Sec.~\ref{sec:gghz-treatment}. We will compare differential
distributions obtained with the \HZNNLOPS{} code and \MCFM{}. The
plots presented in this appendix show the result obtained at the level
of Les Houches events before interfacing with parton shower. We will
also comment on the differences with respect to the results after
parton showering, shown in Sec.~\ref{sec:pheno}.

\begin{figure}
  \centering
  \includegraphics[page=1,width=205pt]{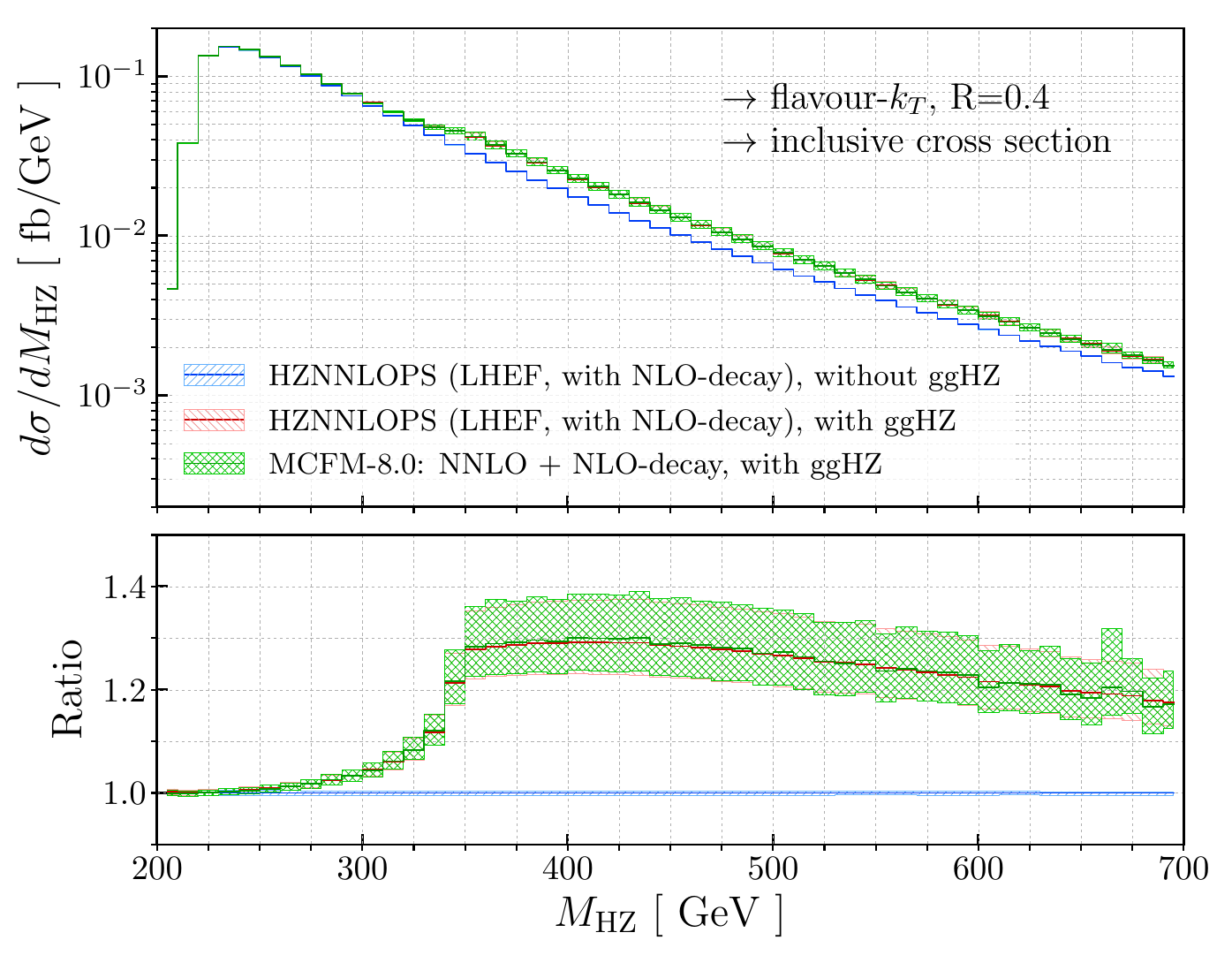}
  \hspace*{0.25cm}
  \includegraphics[page=1,width=205pt]{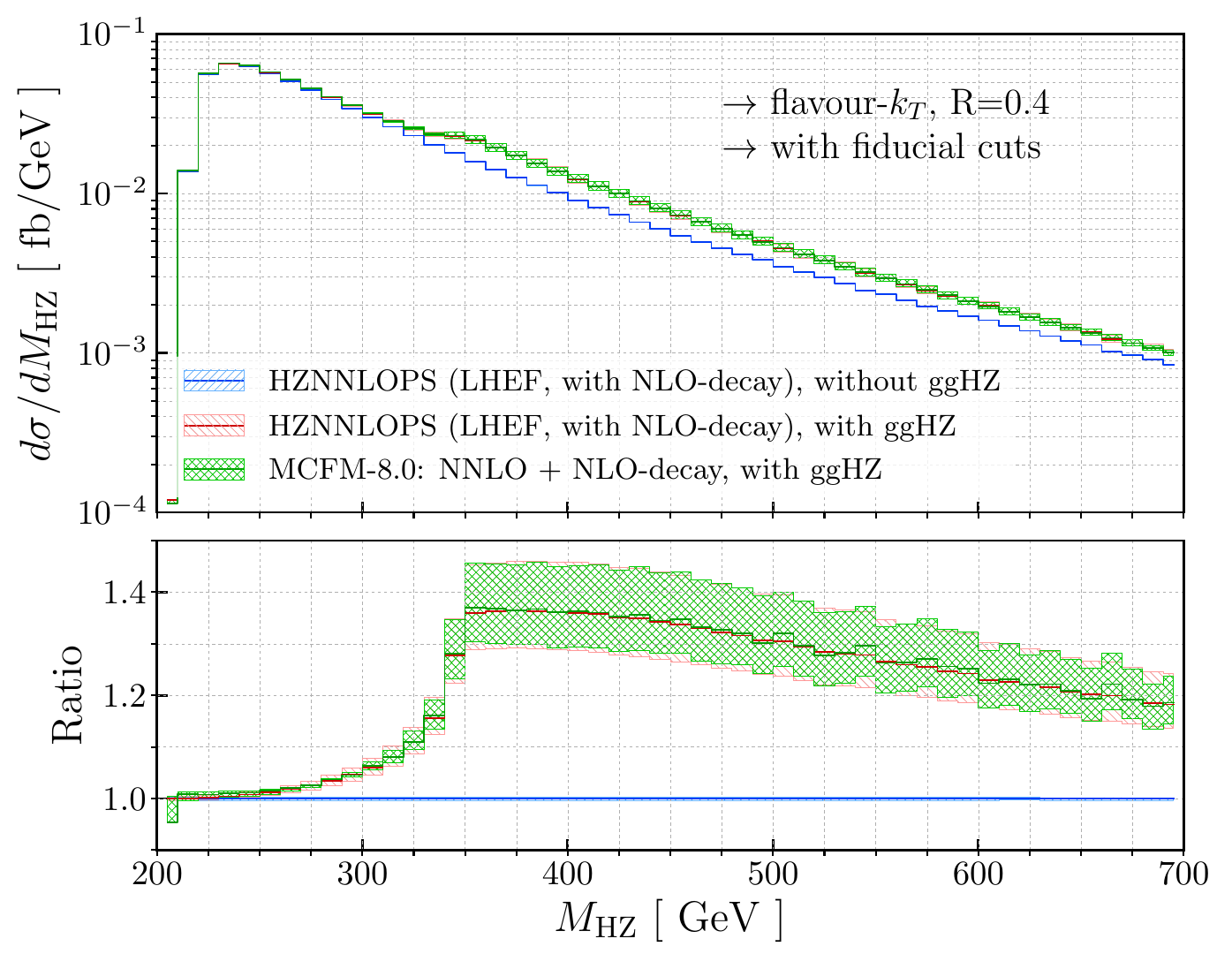}
  \caption{The differential distributions of the invariant mass of the
    \HZ{} system at the LHEF level, without (left panel) and with
    fiducial cuts (right panel).}
  \label{fig:mhz-app}
\end{figure}

The effect of the $\ggHZ{}$ contribution on the invariant mass of the \HZ{} system is shown in Fig.~\ref{fig:mhz-app}, with the left and right panels showing the inclusive and fiducial distributions respectively.
The distributions clearly show that, at LHEF level, the full
\HZNNLOPS{} result matches with \MCFM{}, validating our procedure. One
can also see that the $\ggHZ{}$ contribution is significant when
$\mhz$ is close to, or larger than, the $t\bar{t}$
threshold, and that the corrections remain large well above
threshold. The application of fiducial cuts results in a further
increase of the size of the corrections which is due to differences in
acceptance rates for the two contributions (more than $60\%$ $\ggHZ{}$
events pass the cuts compared to $\sim 45\%$ \HZNNLOLHE{} events).

\begin{figure}
  \centering
  \includegraphics[page=1,width=205pt]{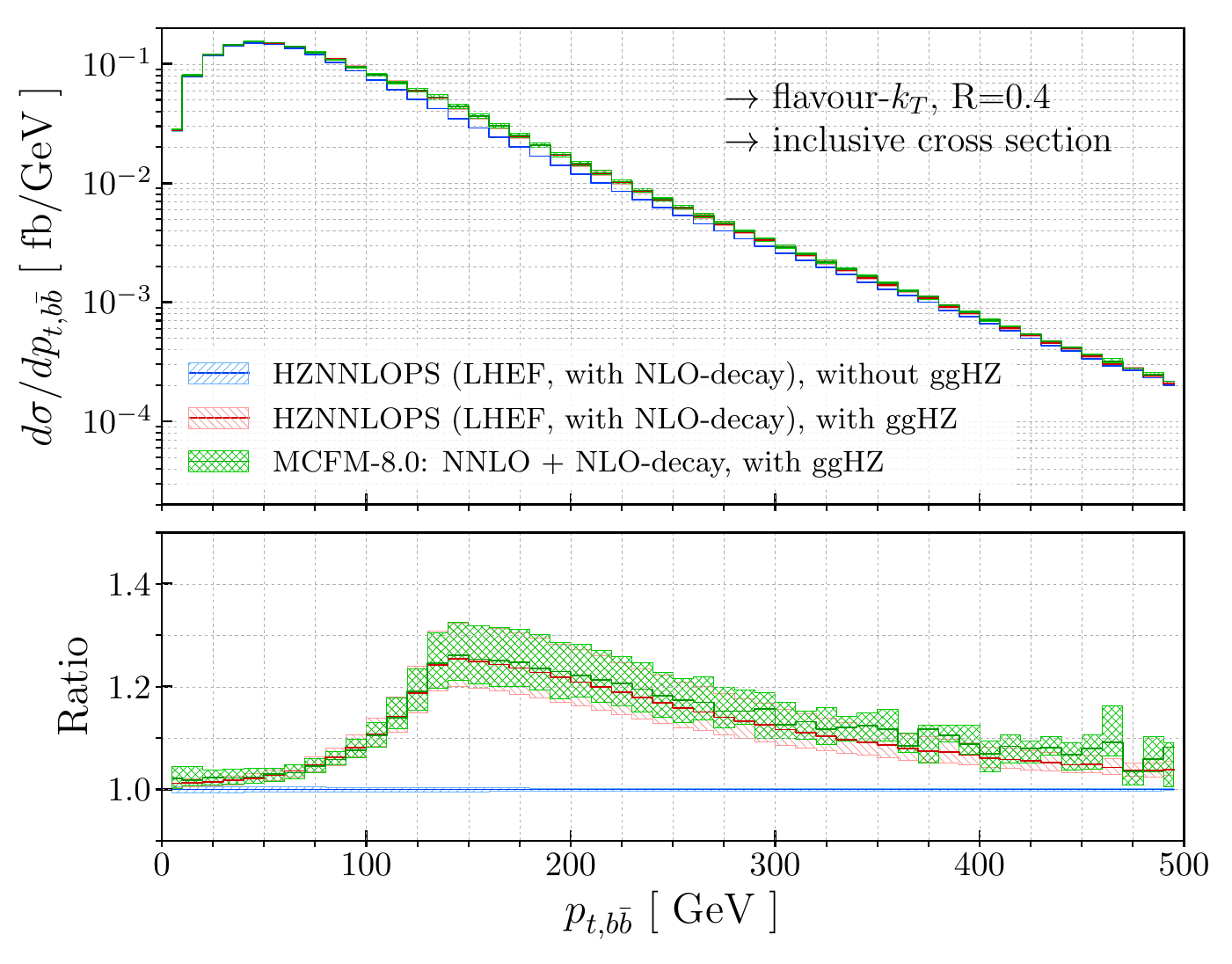}
  \hspace*{0.25cm}
  \includegraphics[page=1,width=205pt]{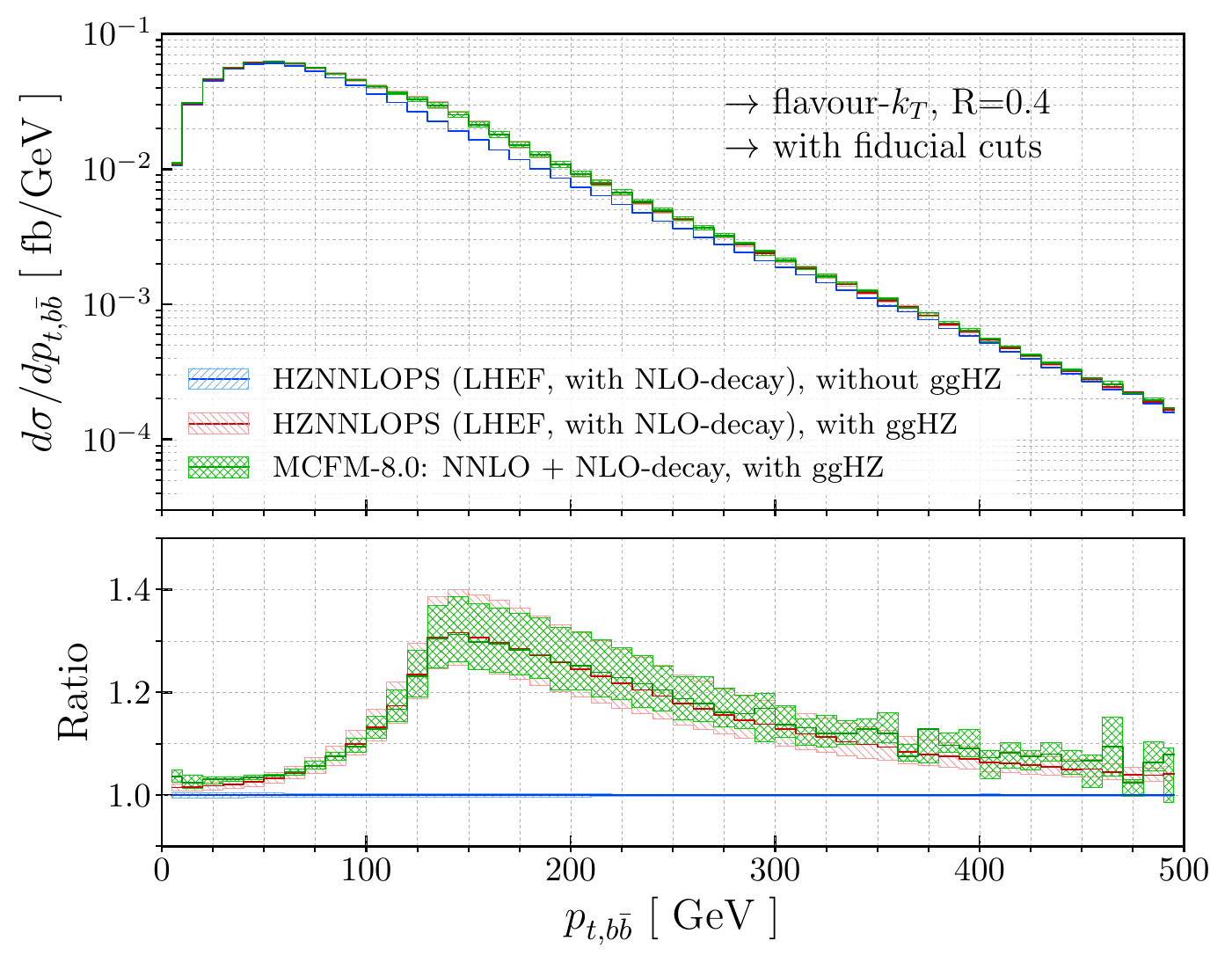}
  \caption{The differential distributions of the transverse momentum
    of the Higgs boson at the LHEF level, without (left panel) and
    with fiducial cuts (right panel).}
  \label{fig:pth-app}
\end{figure}

Similar considerations apply to the distribution of transverse
momentum of the Higgs boson reconstructed from two $b$-jets whose
invariant mass is closest to the Higgs mass. These distributions are
shown in Fig.~\ref{fig:pth-app}. One noticeable difference is that the
effect of the $\ggHZ{}$ dies out faster at large transverse momentum.

\bibliography{hz_nnlops}{}
\bibliographystyle{JHEP}

\end{document}